\documentclass[jgrga]{agutex}

\newcommand\gsim{\,\lower3pt\hbox{$\sim$}\llap{\raise2pt\hbox{$>$}}\,}
\newcommand\lsim{\,\lower3pt\hbox{$\sim$}\llap{\raise2pt\hbox{$<$}}\,}

\usepackage{graphicx}
\usepackage{longtable} 
\usepackage{lscape}
\usepackage{url}
\usepackage{epsfig}
\usepackage{rotating}

\authorrunninghead{LUGAZ ET AL.}

\titlerunninghead{SHOCKS INSIDE CMES: 1997--2006}

\authoraddr{Corresponding author: N. Lugaz,
Department of Physics and Space Science Center, University of
New Hampshire, Morse Hall, 8 College Rd, Durham, NH 03824, USA.
(noe.lugaz@unh.edu)}

\begin{document}

%
%

\title{Shocks inside CMEs: A Survey of Properties from 1997 to 2006}

\authors{N.\ Lugaz\altaffilmark{1}, C.~J.~Farrugia\altaffilmark{1}, C.~W.~Smith\altaffilmark{1}, K.~Paulson\altaffilmark{1}}

\altaffiltext{1}{Space Science Center and Department of Physics, University of New Hampshire, Durham, NH, USA}

%
%

\begin{abstract}
We report on 49 fast-mode forward shocks propagating inside coronal mass ejections (CMEs) as measured by Wind and ACE at 1~AU from 1997 to 2006. Compared to typical CME-driven shocks, these shocks propagate in different upstream conditions, where the median upstream Alfv{\'e}n speed is 85~km\,s$^{-1}$, the  proton $\beta = 0.08$ and the magnetic field strength is 8~nT. These shocks are fast with a median speed of 590~km\,s$^{-1}$ but weak with a median Alfv{\'e}nic  Mach number of 1.9. They typically compress the magnetic field and density by a factor of 2-3. The most extreme upstream conditions found were a fast magnetosonic speed of 230~km\,s$^{-1}$, a plasma $\beta$ of 0.02, upstream solar wind speed of 740~km\,s$^{-1}$ and density of 0.5 cm$^{-3}$. Nineteen of these complex events were associated with an intense geomagnetic storm (peak Dst under $-100$~nT) within 12 hours of the shock detection at Wind, and fifteen were associated with a drop of the storm-time Dst index of more than 50 nT between 3 and 9 hours after shock detection. We also compare them to a sample of 45 shocks propagating in more typical upstream conditions. We show the average property of these shocks through a superposed epoch analysis, and we present some analytical considerations regarding the compression ratios of shocks in low $\beta$ regimes. As most of these shocks are measured in the back half of a CME, we conclude that about half the shocks may not remain fast-mode shocks as they propagate through an entire CME due to the large upstream and magnetosonic speeds.  
\end{abstract}

\begin{article}

\section{INTRODUCTION} \label{intro}
Fast magnetosonic forward shocks are measured at Earth preceding fast coronal mass ejections (CMEs) and, more rarely, propagating into the slower stream of corotating interaction regions (CIRs). Many shocks are also observed without clear drivers, which probably reflects the fact that the angular extent of shocks is larger than that of the putative, associated CME. Hereafter, we use CME to refer to coronal mass ejections, whether when they are observed remotely, in the corona and in the heliosphere, or  when they are measured {\it in situ}. During solar cycle 23, 250--300 shocks were measured by ACE and Wind, just upstream of Earth \citep[e.g.][]{CWang:2010, Janvier:2014}. Interplanetary shocks are followed by a sudden storm commencement (SSC) or a sudden impulse (SI), where the magnetic field strength of low-latitude ground magnetometers suffers a sudden increase, typically associated with a compression of the magnetosphere by the solar wind dynamic pressure, and an increase in the Dst index. The sheath region behind a shock may be geo-effective; in fact, about 15$\%$ of intense geomagnetic storms (as defined by Dst$_\mathrm{peak} < -100$~nT) in solar cycle 23 were caused by CME sheaths \citep[e.g., see][and references therein]{Zhang:2007}. Typical slow solar wind conditions at 1~AU are Alfv{\'e}nic speed of $\sim 50$~km\,s$^{-1}$, density of 5 cm$^{-3}$ and solar wind speed of 400~km\,s$^{-1}$ \citep[see, for example][]{Schwenn:2006, Ebert:2009}. Most interplanetary shocks at 1~AU are relatively weak with a Mach number in the range of 2--4 and compression ratio of the order of 2 \citep[e.g., see][]{Berdichevsky:2000, Oh:2007} for CME-driven shocks. 
In a survey of 105 quasi-perpendicular shocks, \citet{Richardson:2010b} (thereafter, RC2010) found that about 30\% of such shocks were propagating through or closely following a CME, representing a fraction 2--3 times more important than what can be expected from the proportion of CMEs in the solar wind \citep[which is about 10--15\%, as found in][]{Richardson:2000, Zhao:2009}.

The fact that shocks can propagate through CMEs has been known for a long time, with detections reported in \citet{Ivanov:1982} and \citet{Burlaga:1987} for complex disturbances measured in August 1972 and April 1979, respectively. \citet{Wang:2003a} described in detail two occurrences during solar cycle 23 associated with intense geomagnetic storms. The authors found evidence that the compressed magnetic field of the CME contributed to the enhanced geo-effectiveness \citep[]{Wang:2003c}. Simulations of shocks propagating inside a flux rope or magnetic cloud have been performed by \citet{Vandas:1997} and \citet{Xiong:2006}, among others. \citet{Lugaz:2005b} performed a numerical simulation of the interaction of two CMEs using three-dimensional magneto-hydrodynamic modeling, and studied in detail the propagation of the shock  driven by the second CME inside the first magnetic ejecta, including the variation of the upstream conditions, the shock speed and compression ratio (their Figure 2). They found that, in the main part of the cloud, where $\beta \sim 0.1$, the Alfv{\'e}n speed was $\sim 250$~km\,s$^{-1}$ and the sound speed was $\sim 70$~km\,s$^{-1}$.  Consequently, the shock compression ratio was found to decrease from about 3 before the CME and to reach 1.5 $\pm 0.2$ inside the CME. Based on the low compression ratio and the fact that magnetic ejecta are commonly regions of low density, it is unlikely that shocks inside CMEs may be imaged remotely by wide-angle white-light imagers such as the HIs onboard STEREO. In addition, remote-sensing does not give enough information about the upstream and downstream conditions to determine the shock properties and characteristics. A recent example of a shock propagating inside a CME as measured {\it in situ} was reported for the current solar cycle by \citet{Liu:2014b}.

Here, we expand on the work from RC2010 by studying in greater detail the occurences of shocks propagating through a CME during solar cycle 23 (1997--2008). We do so by combining the list of CMEs measured by Wind of \citet{Richardson:2010} with the list of shocks measured by Wind as identified and analyzed by J. Kasper and M. Stevens at the Center for Astrophysics (CfA) and the list of shocks measured by ACE as identified at the University of New Hampshire (UNH) by researchers and graduate students there \citep[see, for example][]{Paulson:2012}. Our goal in this endeavor is threefold: (1) to finally dispose of the misconception that shocks do not propagate through CMEs, (2) to study the additional geo-effectiveness brought about by the shock's compression of the CMEs, and (3) to study shocks in a regime which is unique to CMEs. To build upon this last point, we note that studying shocks inside CMEs give us a unique opportunity to have direct measurements of fast shocks in a magnetically dominated plasma (where the proton $\beta$ is typically less than 0.1) and where the Alfv{\'e}n speed may reach values of several hundreds of km\,s$^{-1}$. 
In section \ref{sec:examples}, we present various examples of shocks propagating through CMEs. In section \ref{sec:stats}, we present an overview of the properties of the shocks, the upstream conditions they propagate into, as well as the geomagnetic effects attributed to the compression of the CME. We also present a superposed epoch analysis of the 49 shocks. We discuss analytical considerations regarding the compression ratio of MHD shocks in low-$\beta$ plasma in section~\ref{sec:analytical}, and discuss our results and conclude in section~\ref{sec:conclusion}.

\begin{figure}[t]
\centering
{\includegraphics*[width=9cm]{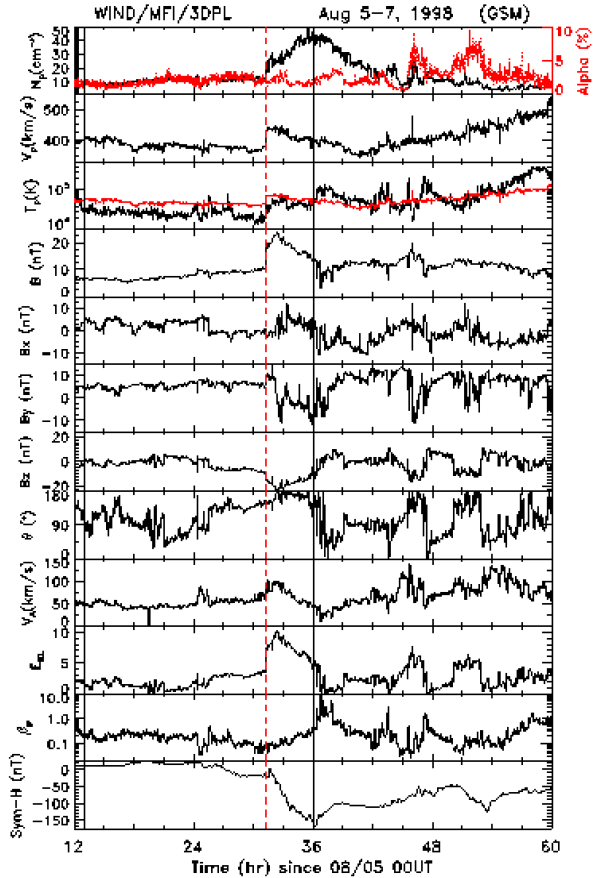}}\\
\caption{Wind observations of the shock inside a CME on 6 August 1998. The slow CME starts around 13:00UT on 08/05 and the shock passes at 07:16UT on 08/06 (31.26 h). The panels show from top to bottom, the proton density (in red the alpha-to-proton number density ratio), the proton velocity and temperature (in red, the expected proton temperature), the total magnetic field and its components in GSM coordinates, the magnetic field clock angle, the Alfv{\'e}n speed, the Kan-Lee coupling function, the proton $\beta$ and the Sym-H index. The CME boundaries as defined by \citet{Richardson:2010} are shown with black lines, the shock with a red dashed line.}
\label{fig:Aug1998}
\end{figure}

\section{Examples of Shocks Inside CMEs}\label{sec:examples}
\subsection{Typical Case: August 6, 1998 shock}
We first present a typical case, chosen so as to be as close as possible to the median from our sample. Figure~\ref{fig:Aug1998} shows from top to bottom, the proton density, speed and temperature, the magnetic field strength, its components and clock angle ($\theta$), the Alfv{\'e}n speed, the Kan - Lee coupling function \citep[]{Kan:1979}, the proton $\beta$, and the Sym-H index. In the proton temperature panel, the red line shows the expected temperature following \citet{Lopez:1987}. The clock angle, $\theta$, is the polar angle in the GSM $YZ$ plane. The Alfv{\'e}n speed takes into consideration the influence of the alpha particles.  The Kan-Lee electric field is defined as: $E_{KL} = V \times B_T \, sin^2(\theta/2)$, where $B_T = (B_y^2 + B_z^2)^{1/2}$, the field components being measured  in GSM coordinates, and $\theta$ is the clock angle \citep[see also][]{Shepherd:2007}.

There is a CME identified by \citet{Richardson:2010} starting around 13:00 UT on August 5, 1998 and lasting until 12 UT, on August 6. It is a relatively slow structure, with an average speed of $\sim 380$~km\,s$^{-1}$ and it does not drive a shock ahead of it. The CME does not fit the typical characteristics of a magnetic cloud \citep[MC, see][]{Burlaga:1981} but it is a relatively clear magnetic ejecta. It is given a subjective ``mark'' of 1 in \citet{Richardson:2010}, where 2 represents a clear MC, and 0 a very weak and irregular ejecta. It is characterized by a lower-than-expected proton temperature, enhanced magnetic field strength and relatively smooth rotation of the magnetic field vector, among other properties. It has however a relatively large density ($\sim 10$~cm$^{-3}$), and therefore, the proton $\beta$, while lower than 1 is only lower than 0.1 in a small part of the event. At 07:16 UT on August 6, Wind observes a fast forward shock, marked by a red vertical guideline on Figure~\ref{fig:Aug1998},  propagating inside this CME. The end boundary of the CME is relatively clear around 12:00 UT on August 6 and corresponds to a short interval of very low magnetic field strength and enhanced temperature and $\beta$, very similar to the interaction region between two CMEs as described in \citet{Wang:2003}.  
Comparing the magnetic field orientations upstream and downstream of the shock, there is almost no change in the clock angle. Overall, these provide strong indications that the shock is, indeed, propagating through the CME and is not simply at the back boundary, which occurs hours behind the shock. The shock occurs about 18.5 hours into the CME and about 5 hours before its end. The source of the shock is unclear: the rising speed on August 7 indicates that a fast solar wind speed is catching up to the CME. However, the time period around 12 UT on August 6 is highly reminiscent of the CME-CME interaction region as described by \citet{Wang:2003}. 

The conditions upstream of the shock are $N_p \sim 11$~cm$^{-3}$, $V_p \sim$ 370~km\,s$^{-1}$, B $\sim$ 10 nT and $T_p \sim 10^4$~K, which correspond to $V_a \sim 70$~km\,s$^{-1}$ and $\beta \sim 0.07$. The shock is quasi-perpendicular, with an angle $\Theta_{Bn} = 80^\circ \pm 5^\circ$. The density and magnetic field compression ratio are 1.8 $\pm 0.2$, the shock speed is 460 $\pm 40$~km\,s$^{-1}$, which corresponds to a fast Mach number of 1.5 $\pm 0.3$. The $B_z$ component of the magnetic field decreases from $-6$ nT to $-11$ nT through the shock and reaches a peak at $-20$~nT about 1.3 hours behind the shock. The increase in velocity and magnetic field strength behind the shock results in a near tripling of the coupling function to $E_\mathrm{KL} = 10$ mV~m$^{-1}$, values characteristic of strong driving of the magnetosphere by the solar wind. The Dst decreases from $-25$~nT at 07:00 UT (31 h) on August 6 before the shock arrival to $-138$~nT at 12:00 UT (36 h) on the same day. The Sym-H index (equivalent to Dst but with a 1-minute resolution), which is what is plotted in Figure~1, drops from $-16$~nT at 07:15 UT on August 6 to $-169$~nT at 12:14 UT on the same day. This intense geomagnetic storm is quite clearly associated with the compression of the magnetic field and the increased velocity behind the shock. Indeed, the period between the shock and the end of the CME is the only time when the coupling electric field, $E_\mathrm{KL}$, reaches high values ($>$5 mV\,m$^{-1}$). So the major geomagnetic effects are related to this shock. In fact, the Sym-H decrease, the enhancement of magnetosphere convection and two substorms (not shown here) all occur in the time period from 8 to 13 UT on August  6.

This example is representative of our sample of 49 shocks inside a CME, as it occurs about 20 hours after the start of the magnetic ejecta and propagates into a medium with Alfv{\'e}n speed of the order of 80~km\,s$^{-1}$ and a plasma $\beta$ of the order of 0.1, conditions representative of a relatively weak CME. Next, we describe some of the more extreme cases in our sample.

\subsection{Very low Mach number shock: November 8, 1998 }
\begin{figure}[t]
\centering
{\includegraphics*[width=9cm]{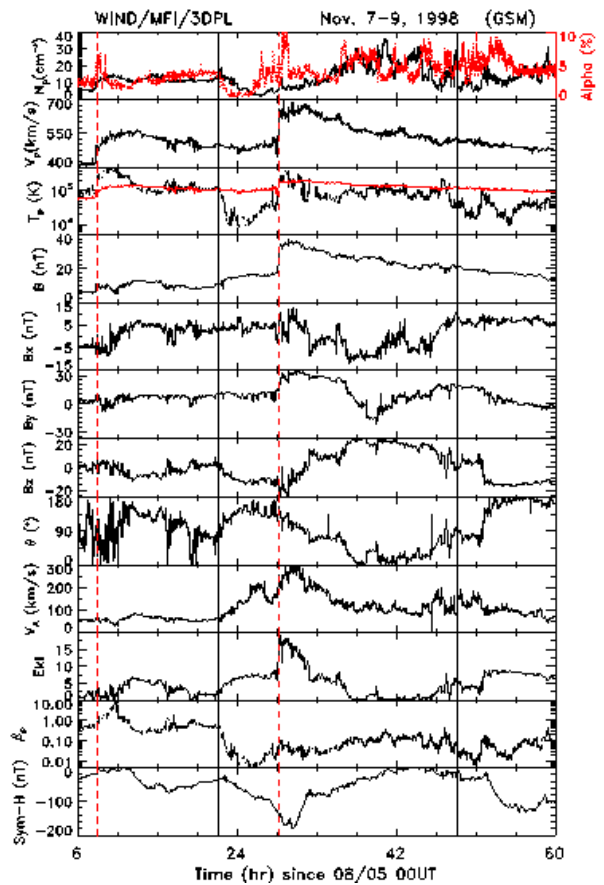}}\\
\caption{Wind observations of the shock inside a CME on 11/07/1998. The CME starts around 22:00UT on 11/07 and is preceded by a shock at 08:15UT on the same day (marked with the first dashed red line). A second shock passes at 04:51UT on 11/08 (28.85 h), and is marked by the second red dashed line.}
\label{fig:Nov1998}
\end{figure}

Figure~\ref{fig:Nov1998} shows Wind measurements around the passage at Earth of a shock inside a CME on November 8, 1998. This Figure follows the same format as Figure~1. A  shock passed Wind at 08:15 UT on November 7 followed by a MC-like ejecta starting around 22:00 UT (between the black vertical guidelines). Starting at that time, the temperature is lower than the expected one and the proton $\beta$ is $\sim  0.1$, which are conditions typical of the passage over the spacecraft of a magnetic ejecta. At 04:51 UT on November 8, there is a fast forward shock propagating through this CME. The upstream conditions in the 10 minutes preceding the shock passage are: $N_p \sim 4.5$~cm$^{-3}$, $V_p \sim$ 450~km\,s$^{-1}$, B $\sim$ 16.5 nT and $T_p \sim 4 \times 10^4$~K, which correspond to $V_a \sim 170$~km\,s$^{-1}$ and $\beta \sim 0.03$. The shock occurs about 6 hours after the start of the magnetic ejecta (as defined by \citet{Richardson:2010}). It is oblique, with an angle $\Theta_{Bn} = 65^\circ \pm 10^\circ$. The density and magnetic field compression ratios are  2 $\pm 0.2$, the shock speed is 600 $\pm 50$~km\,s$^{-1}$, which corresponds to a fast magnetosonic Mach number of 1.4 $\pm 0.3$. The $B_z$ component of the magnetic field decreases to $-25$ nT after the shock passage but remains enhanced as compared to the upstream for less than 2 hours. The main change in the magnetic field is a large increase in the $B_y$ component. The Sym-H index decreases after the arrival of the shock to reach two minima 45 minutes and 1.5 hours after the shock passage with minimum Sym-H values of $-173$ and $-180$~nT. This corresponds to a decrease of  Sym-H by about 40~nT from its pre-shock value. Sym-H  was decreasing before the shock passage and it is impossible to quantify the effects of the shock alone. However, the second dip closely following the first dip is certainly due to the shock effects. 

Note that, even for such a low Mach number, the shock compresses the magnetic field and density by a factor of $\sim 2$ and may significantly affect the geo-effectiveness of the CME. As for the previous example, the magnetic field clock angle does not vary much through the shock. The enhancement of density downstream of the shock is very short-lived, which is a common occurrence for these shocks with very low Mach number (it happens for about 25\% of the shocks in our sample). However, the increases in the magnetic field and pressure (or temperature) are long-lasting. This shock has the second strongest upstream magnetic field and is the one occurring closest to the CME leading edge (as a percentage of the total CME duration). 

A further example with low Mach numbers is the January 1, 2006 shock which propagates into the following conditions: $N_p \sim 5$~cm$^{-3}$, $V_p \sim$ 440~km\,s$^{-1}$, B $\sim$ 10 nT and $T_p \sim 5 10^4$~K, which correspond to $V_a \sim 90$~km\,s$^{-1}$ and $\beta \sim 0.08$. The shock is quasi-perpendicular ($\Theta_{Bn} = 85^\circ \pm 5^\circ$) with a speed of 450 $\pm 50$~km\,s$^{-1}$, which gives a Mach number of 1.2 $\pm 0.3$. The compression ratio for the density and magnetic field are 1.6 $\pm 0.2$. As the shock occurs during a northward-$B_z$ period, there is no associated geo-effectiveness.

\subsection{Fast Upstream and Alfv\'en Velocities, low plasma $\beta$: September 12, 2005}
\begin{figure}[t]
\centering
{\includegraphics*[width=9cm]{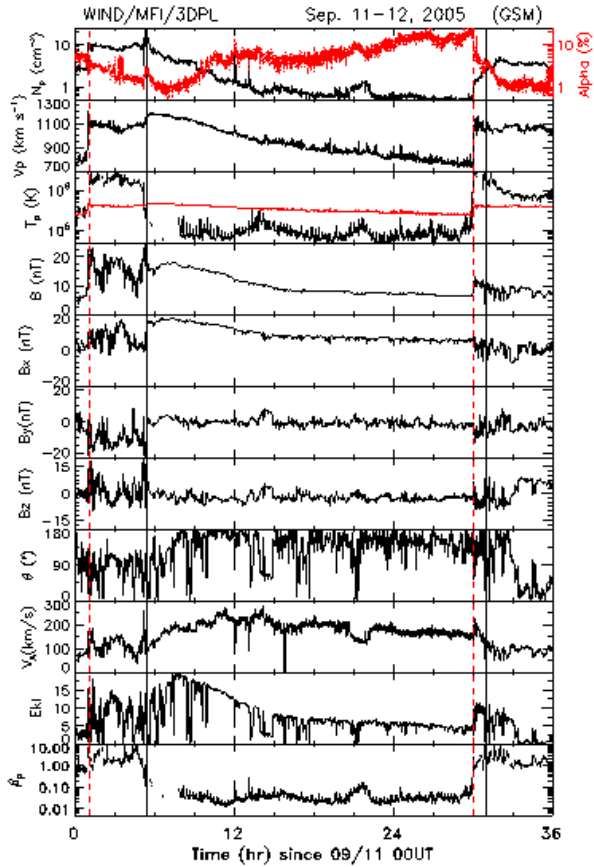}}\\
\caption{Wind observations of the shock inside a CME on 09/12/2005. The CME starts around 05:00UT on 09/11 and is preceded by a shock at 01:14UT on the same day. A second shock passes at 06:00UT on 09/12 (30h).}
\label{fig:Sep2005}
\end{figure}

We next present a case with the largest upstream solar wind velocity $V = 740$~km\,s$^{-1}$, which also corresponds to one of the largest upstream Alfv{\'e}n speeds. Figure~\ref{fig:Sep2005} shows the plasma and magnetic field measurements by Wind.
There is a shock on 01:14 UT on September 11, 2005 which precedes a fast CME. The CME duration, as reported by \citet{Richardson:2010} is 26 hours (from 05 UT, September 11 to 07 UT, September 12), and it is characterized by high magnetic field strength, very low density, low plasma $\beta$, decreasing speed profile. The magnetic field components are not particularly smooth, and this event is marked as 0 in \citet{Richardson:2010}. At 06:00 on September 12, Wind encounters a fast-mode forward shock. It occurs at or close to the back boundary of the CME, and 25 hours after its start. At that time, the magnetic field is mostly in the radial direction, and in addition to the large speed and very low density, this leads us to consider that the shock is propagating through the rarefaction region behind a CME. Development of such rarefaction regions is often seen in numerical simulations \citep[e.g., see][]{Manchester:2014} and can also be inferred from remote heliospheric observations \citep[]{THoward:2013b}. 

The conditions upstream of the shock are $N_p \sim 0.5$~cm$^{-3}$, $V_p \sim$ 740~km\,s$^{-1}$, B $\sim$ 6.5 nT and $T_p \sim 9 \times 10^4$~K, which correspond to $V_a \sim 200$~km\,s$^{-1}$ and $\beta \sim 0.04$. The shock is quasi-parallel, with an angle $\Theta_{Bn} = 20^\circ \pm 5^\circ$. The density and magnetic field compression ratio are  2 $\pm 0.2$, the shock speed is 1000 $\pm 50$~km\,s$^{-1}$, which corresponds to a fast Mach number of 1.5 $\pm 0.3$. The $B_z$ component of the magnetic field decreases from $-2$ nT to $-10$ nT shortly after the shock but does not remain southward, consequently there are little changes in the Dst index after the shock.

\section{Upstream Conditions and Properties of Shocks Inside CMEs}\label{sec:stats}

\subsection{Data Selection and List}
We started from the list of CMEs of \citet{Richardson:2010} measured by Wind and compared it with the list of shocks measured by Wind as identified and analyzed at CfA and the list of shocks measured by ACE as identified at UNH by researchers and graduate students there. We identified 59 shocks propagating inside or at the back of CMEs. {\it In situ} measurements were visually analyzed for each of these cases, which made us remove 10 cases for which the upstream conditions were close to typical, or for which the perturbation was not a shock. We ended up with 49 shocks propagating through CMEs.
Our list contains all events identified by RC2010 to be shocks propagating inside a CME or a CME trailing edge except the following six exceptions: one shock propagating inside a potential CME (June 14, 2005 marked as CME?), for which we do not consider the upstream medium to be a CME, two shock propagating through CME trailing edges (February 11, 2000 and March 31, 2001) for which the upstream conditions are close to that of the typical solar wind, and we therefore removed from the sample, and three shocks identified by RC2010 as propagating inside CMEs (May 4, 1998, May 23, 2000 and October 25, 2001). For the first two of these three shocks, it was not clear that they satisfy the Rankine-Hugoniot relations for fast-mode forward shocks. The last case had upstream conditions comparable to the solar wind, and we also removed it from our sample. 

Table~1 lists the 49 shocks propagating through CMEs, with the columns showing, from left to right, the CME start and end times and type as listed by \citet{Richardson:2010}, the shock time at Wind, the delay since the start of the CME in hrs, the shock speed in the Earth's rest frame, the shock angle, density and magnetic field compression ratios, the shock Alfv{\'e}nic Mach number, the upstream fast magnetosonic and Alfv{\'e}n speeds, proton $\beta$ and solar wind speed, the minimum Dst reached within 12 hours of the shock detection at Wind, the drop in Dst from the value before the shock arrival to the minimum in the next 12 hours and the time when the minimum Dst was reached. Out of the 49 shocks in our list, 25 were identified in RC2010: 21 were identified to propagate inside a CME, 1 inside a possible CME (marked as CME?), 2 inside a CME trailing edge and 1 inside a CME sheath. We also identified 24 additional events; some are clearly not quasi-perpendicular shocks, whereas others appear to be and it is unclear why they were not in the list of RC2010. There were no shocks inside CMEs in 2007 and 2008. For the Bastille day events, we used the values provided by \citet{Lepping:2001}.  Missing values are marked with a $\_$. 

Taking into account that two CMEs have multiple shocks propagating inside them, and comparing with the full list of CMEs measured at 1 AU from \citet{Richardson:2010}, we find that about one out of every 7 CMEs has a shock propagating through it (47/322). There is no broadly used, dedicated shock list; \citet{CWang:2010}, for example, list 257 well-defined shocks from 1998 to 2008. Using this value for the total number of shocks, we find that about 19\% of all shocks propagate inside CMEs. This is a lower proportion than that reported by RC2010 for quasi-perpendicular shocks ($\sim 30\%$). Further discussion is presented in section \ref{sec:conclusion} about these points. 

In order to compare our dataset with a similar sample of shocks propagating into typical solar wind conditions, we selected 45 shocks during the years 1997--2006 with approximately the same yearly distribution as the sample of shocks inside CMEs. Hereafter, we refer to this sample of shocks as the ``control sample''. These shocks were selected from the list of well-determined shocks measured by ACE (UNH list) and Wind (CfA list). The average properties of this control sample are as follows: the fast magnetosonic Mach is 2.34 (average Alfv\'en Mach of 2.84), the average shock speed is 540~km\,s$^{-1}$, the average $\beta$ is 0.63 (median 0.49), the average compression ratio is 2.55 (median 2.4). This sample has average and median properties close to but slightly higher than that found by other researchers for a larger number of shocks. For example, \citet{Oh:2007}, found a typical compression ratio of 2, based on 249 shocks from 1995 to 2001, an Alfv\'en Mach of 2.1, a typical shock speed of 490 km\,s$^{-1}$, and a upstream $\beta \sim 0.3$.  From the list of \citet{CWang:2010}, the average speed of the 257 shocks is  500~km\,s$^{-1}$, the average fast magnetosonic Mach is 2.06 and the average compression ratio is 2.06. The larger values for our control sample may be due to the fact that it includes more shocks during the rise and solar maximum phases (1999--2002) than a random sample would have. Shocks at solar maximum tend to be faster and have larger Mach numbers and compression ratio \citep[]{Oh:2007}. This may also reflect the fact that this sample excludes shocks inside CMEs which tend to have lower compression ratio and Mach numbers, as discussed below.

\subsection{Superposed Epoch Analysis}

\begin{figure*}[t]
\centering
{\includegraphics*[width=7.cm]{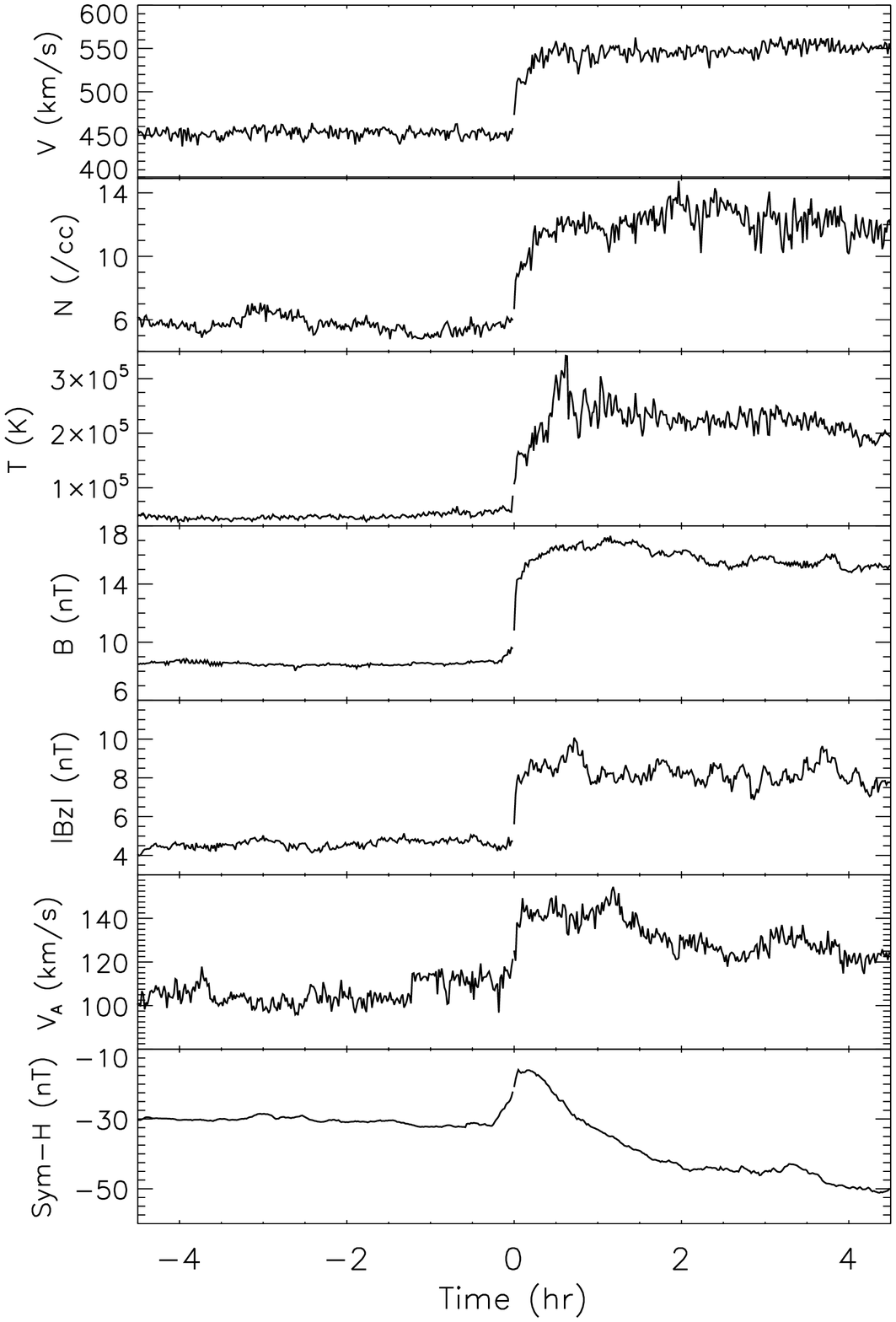}}
{\includegraphics*[width=7.2cm]{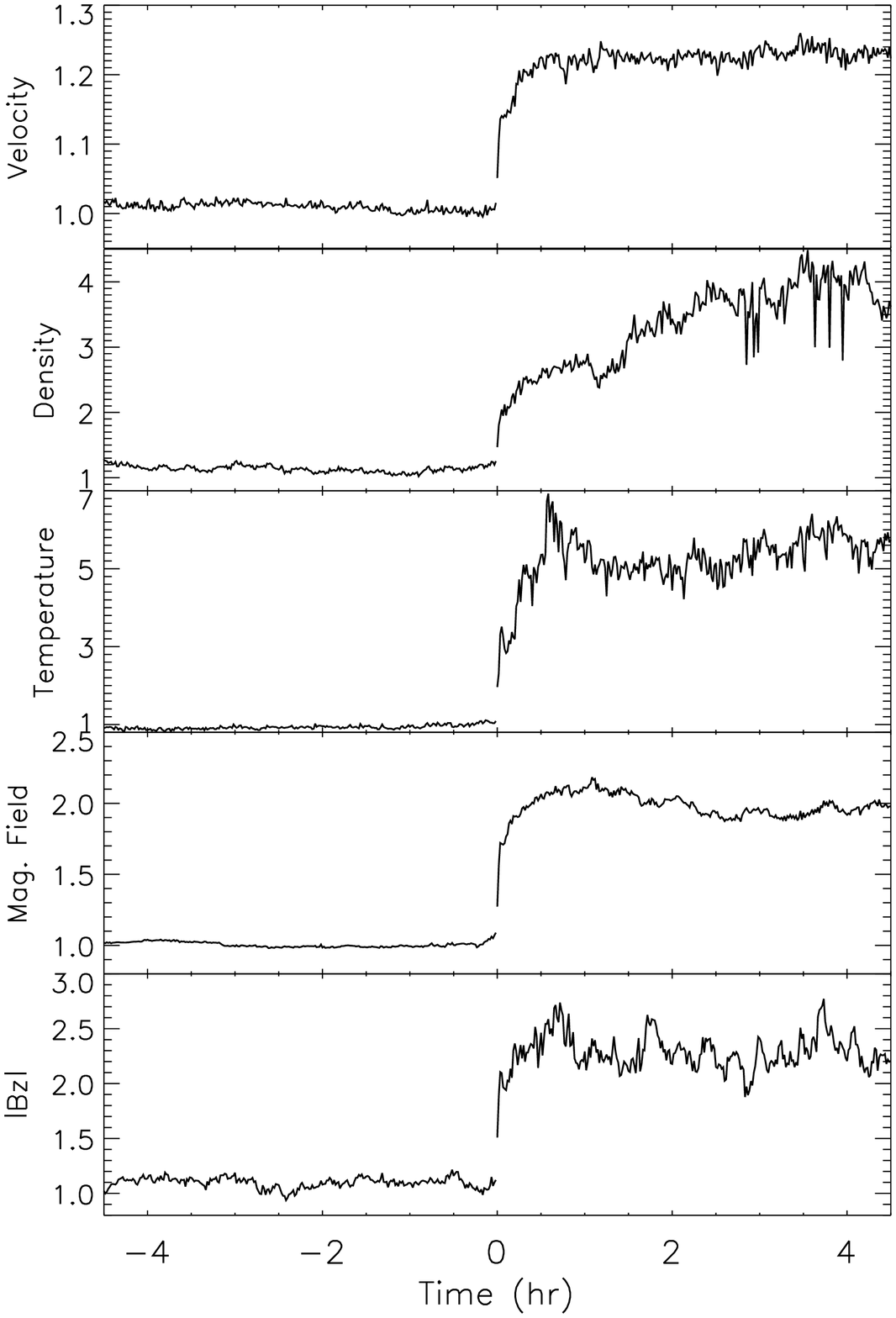}}\\
\caption{Superposed epoch analysis of the 49 shocks with data from omniweb. The left panels show the velocity, proton density, proton temperature, total magnetic field, absolute value of the magnetic field $B_z$ component in GSM coordinates, Alfv{\'e}n speed and Sym-H index, from top to bottom. The right panels show the dimensionless velocity, density, temperature, total and $B_z$ magnetic field, from top to bottom.}
\label{fig:superposed}
\end{figure*}

We performed a superposed epoch analysis of the 49 shocks propagating inside a CME, using 1-minute data from omniweb\footnote{\url{http://omniweb.gsfc.nasa.gov}}.   Omniweb data is time-shifted to the nose of Earth bow shock, providing a common temporal reference point for all the shocks. For the November 6, 2001 event, there was no data after the shock arrival (except Sym-H) and for the July 15, 2000, there was no data except Sym-H in the considered time window. These were associated with the two strongest drops of Sym-H after the shock from our dataset, so we included them nonetheless in our analysis. The October 31, 2001 has about 3.5 hours of data missing around the shock arrival time, and it is also included in the sample. We visually identify the shock arrival time to serve as the zero epoch. Figure~\ref{fig:superposed} shows the velocity, density, temperature, magnetic field strength, absolute value of GSM $B_z$ component, Alfv{\'e}n speed and Sym-H index for 4.5 hours before and after the shock arrival time.  
Because a few events with extreme values may affect the average of the events, we also plot in the right panel of Figure~\ref{fig:superposed}, the velocity, density, magnetic field strength, GSM $B_z$ component and temperature in dimensionless form. To do this, we calculate, for each event, the maximum value of the parameters before the shock arrival and we scale the measurements by that value before doing the averaging. The averaged results are then scaled so that the pre-shock value is about 1. In this way, any event with extreme values of the plasma and magnetic parameters shall have less impact on the global trend. It is also easy to visualize the approximate average jump at the shock with this plot. The same plots are shown for the control sample in the Appendix.

The left panel of Figure~\ref{fig:superposed} clearly shows the average upstream conditions into which the shocks propagate: velocity of 450~km\,s$^{-1}$, density of 6 cm$^{-3}$, magnetic field of 8.5~nT (with a $B_z$ component of about 4.5~nT), and an Alfv{\'e}n speed of about 105~km\,s$^{-1}$. These average values are very close to the median (average) from our sample as listed in Table~1, which are 430 (455)~km\,s$^{-1}$, 4.5 (5) cm$^{-3}$, 8 (8.5) nT, a Alfv{\'e}n speed of 85 (105)~km\,s$^{-1}$ and a plasma $\beta$ of 0.15 (0.08). These upstream conditions differ from typical ones as found for our control sample: upstream velocity of 410~km\,s$^{-1}$, magnetic field of 6 nT and Alfv\'enic speed of 70~km\,s$^{-1}$. A difference of 30-40~km\,s$^{-1}$ in both upstream solar wind and Alfv\'enic speeds may not sound like much, but, for a shock with a speed of 600~km\,s$^{-1}$, the Alfv\'enic Mach number would be 2.7 for normal conditions but only 1.4 for upstream conditions as encountered inside CMEs. 

As shown in the right panel of Figure~\ref{fig:superposed}, the shock compresses, on average, the density by a factor of 2, and the magnetic field by a factor of 1.7. The average increase in the magnitude of the $B_z$ component is about 2. As shown in the left panel of Figure~\ref{fig:superposed}, the magnetic field continues to increase behind the shock, so that the maximum magnetic field strength is about 17.5 ~nT, 1 hour behind the shock, and the average density increases to about 13.5~cm$^{-3}$. The jump through the shock in the velocity is about 100~km\,s$^{-1}$. 

\begin{figure*}[h]
\centering
{\includegraphics*[width=8.cm]{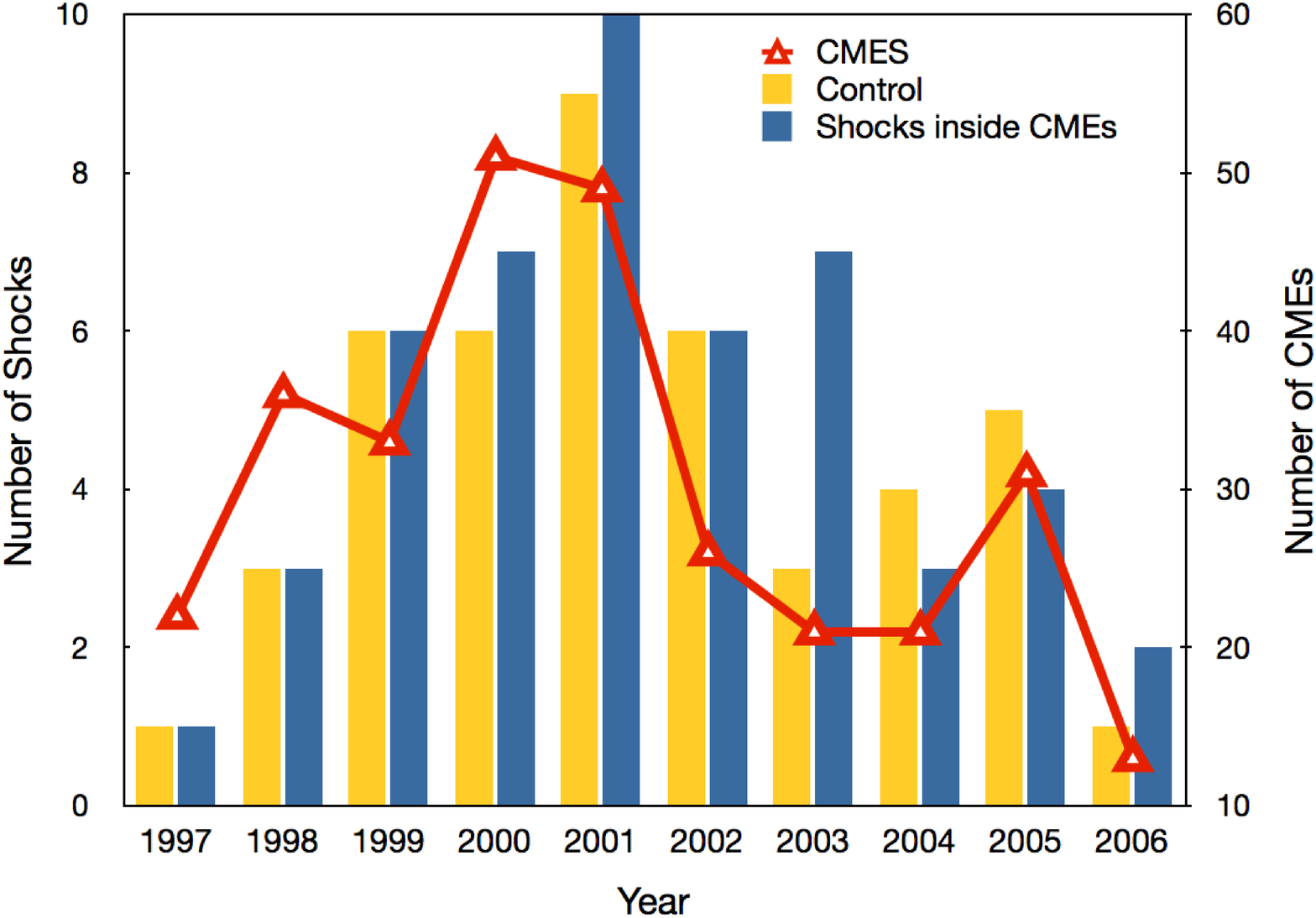}}
{\includegraphics*[width=7.8cm]{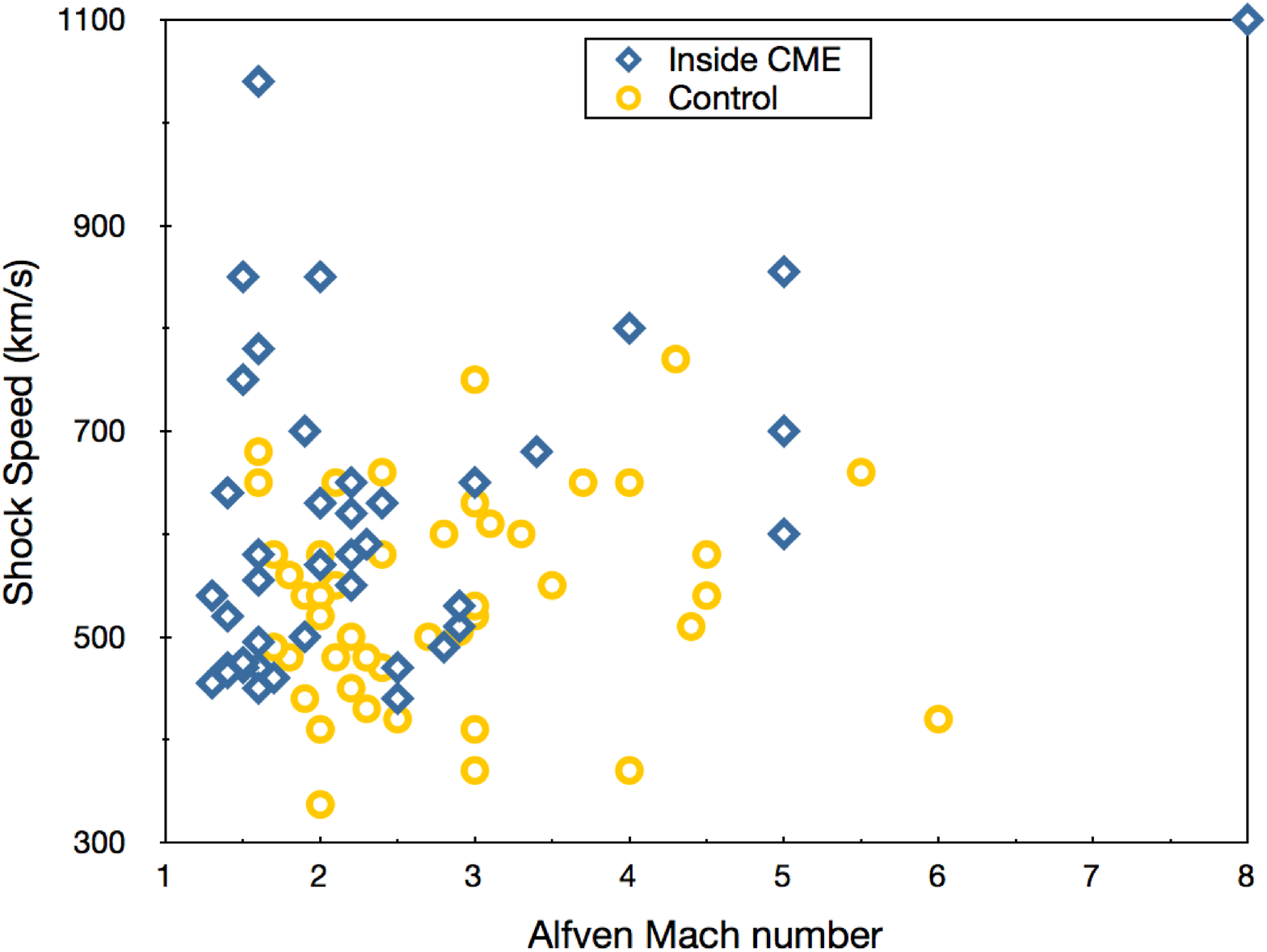}}
\caption{Left panel: yearly distribution of the shocks inside CMEs (blue bars) and shocks from the control sample (yellow bars) as well as total number of CMEs measured {\it in situ} (red triangles). Right panel: Shock speed and Alfv{\'en} Mach number for the shocks inside CMEs (blue diamonds) and shocks from the control sample (yellow circles).}
\label{fig:shock}
\end{figure*}

Although the events were not selected based on their geo-effectiveness, the average Sym-H index is perturbed at a value of $-30$~nT before the shock arrival. This is somewhat consistent with the fact that the upstream conditions correspond to the passage at 1~AU of a CME \citep[see for example][]{Richardson:2010}. On average, there is a sudden storm commencement at the shock passage with an increase of the Sym-H by about 14 nT, followed by a decrease with the Sym-H reaching $-51$~nT 4.5 hours after the shock passage. Note that the inclusion of the two strongest drops for which there is no downstream plasma data does not change the shape of the Sym-H curve and only changes the minimum value by 3~nT. The shape of the Sym-H curve is very similar to that found for the control sample, but it is shifted by $-20$~nT. Therefore, a moderate geomagnetic storm is the average result from the passage at Earth of a shock inside a CME.

\subsection{Detailed Analysis}

We now give some additional details regarding the upstream conditions and the geo-effectiveness of these shocks. The two subsequent sections discuss the shock properties and location within the CME in more detail.

20 of the 49 shocks propagate inside solar wind with Alfv{\'e}nic speed greater than 100~km\,s$^{-1}$. Some of the largest Alfv\'enic speeds are measured for the three shocks during the Halloween 2003 time period (October 26-28) with upstream Alfv\'enic speeds between 180 and 220~km\,s$^{-1}$. The March 20, 2002 shock has upstream Alfv\'enic speed in excess of 400~km\,s$^{-1}$ due to very low upstream density. In fact, 19 shocks propagate through upstream densities lower than 3~cm$^{-3}$, 13 of which also have Alfv\'enic speed greater than 100~km\,s$^{-1}$. 

Only 15 shocks propagate inside magnetic fields stronger than 10~nT, which is the average magnetic field measured at 1~AU inside CMEs \citep[]{Richardson:2010}. 3 of these propagate inside magnetic fields stronger than 15~nT,  with the November 6, 2001 shock the one propagating into the strongest magnetic field ($\sim 20$~nT, although the upstream density is greater than 10~cm~$^{-3}$). In that sense, many of the CMEs into which the shocks propagate are relatively weak. Most of the CMEs are also relatively slow. 8 shocks occur with upstream speeds greater than 550~km\,s$^{-1}$ with the largest upstream speed of 740~km\,s$^{-1}$ as already discussed for the September 12, 2005 shock. It is expected that shocks are neither able to overtake extremely fast CMEs, nor able to propagate through CMEs with such fast upstream conditions.  This might explain why the median upstream speed (430~km\,s$^{-1}$) for the 49 shocks is slightly lower than the median speed of CMEs measured at 1~AU \citep[476 km\,s$^{-1}$, see:][]{Richardson:2010}.

The average (median) Dst measured within 12 hours of the shock arrival at Wind is $-81$ ($-72$)~nT. 19 of the 49 shocks are associated with an intense geomagnetic storm (Dst$_\mathrm{peak} <-100$~nT) within 12 hours of the shock detection at Wind. The two largest storms are the one on July 15, 2000 where Dst drops from $-55$~nT to $-289$~nT in the 8 hours following the shock arrival and the one on November 6, 2001 where the Dst drops from $-73$~nT to $-292$~nT 5.5 hours after the shock arrival. 
15 shocks were associated with a drop of the Dst index of more than $-50$~nT within 9 hours of the shock passage. The average drop within 12 hours is $-46$~nT (median $-32$~nT). Since these shocks occur inside CMEs, it is not always possible to separate exactly the effects of the CME with that of the compression from the shock. 
\begin{figure*}[t]
\centering
{\includegraphics*[width=7.8cm]{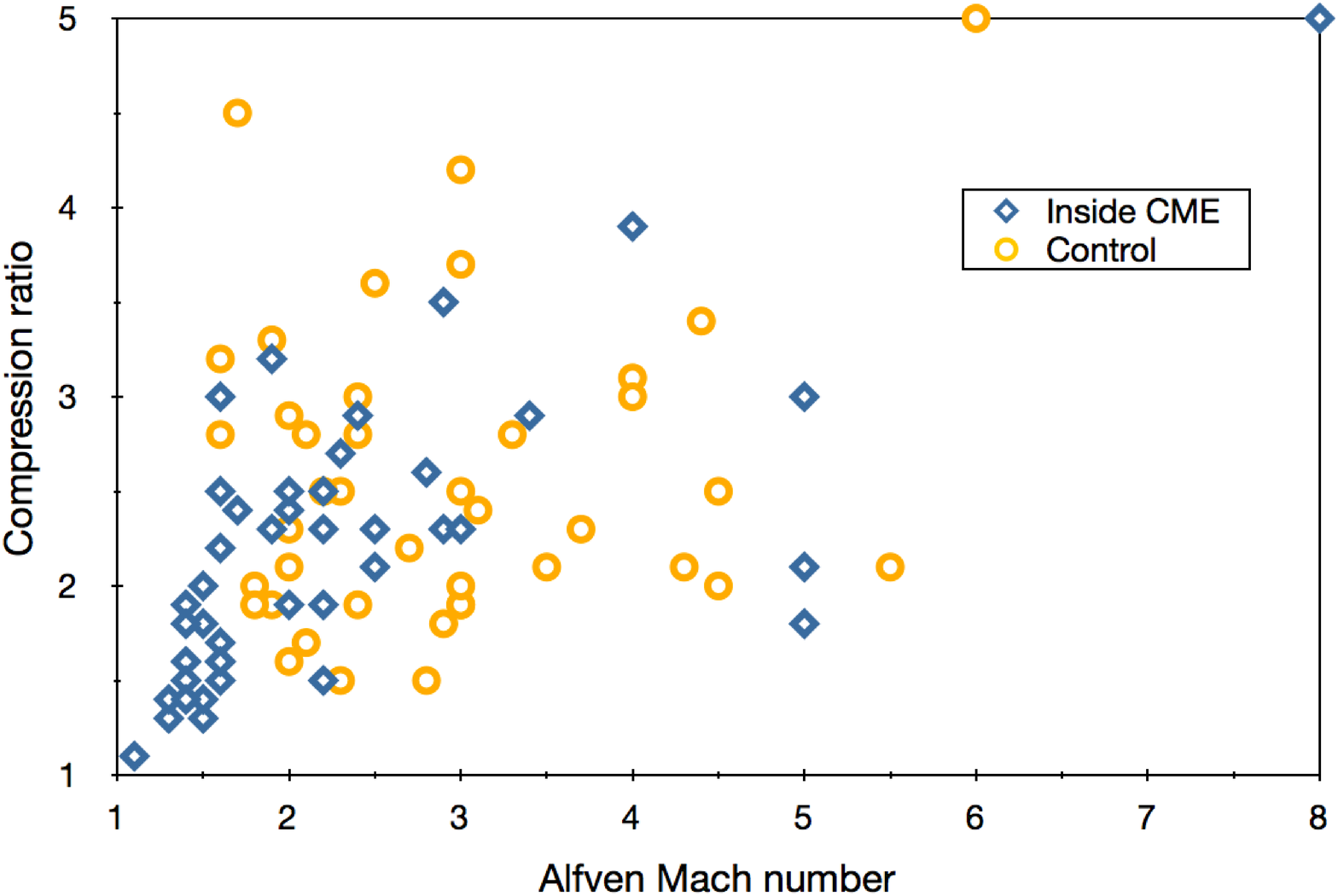}}
{\includegraphics*[width=7.8cm]{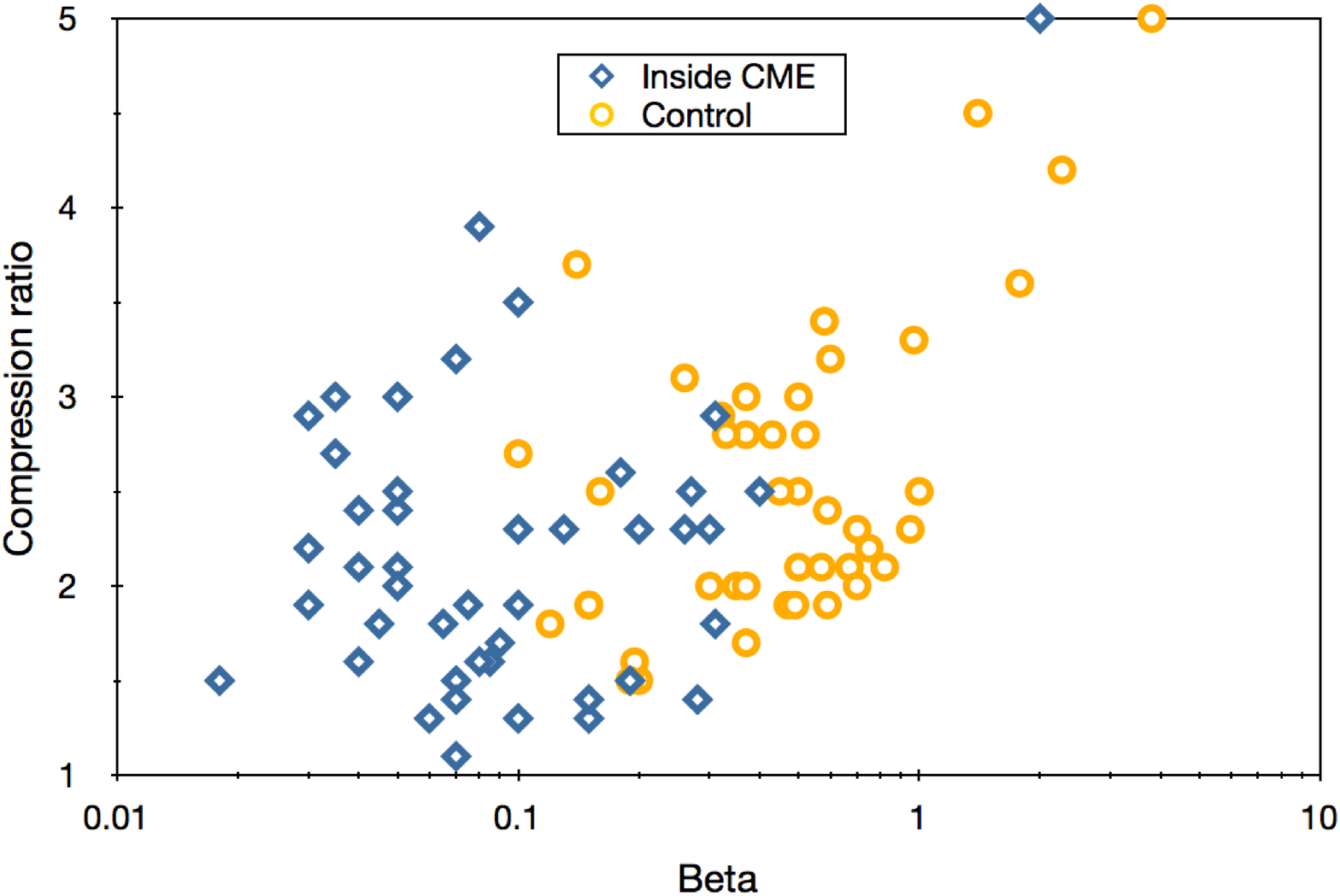}}\\
\caption{Compression ratio of the shocks vs Alfv{\'e}n Mach number (left panel) and plasma $\beta$ (right panel) for the shocks inside CMEs (blue diamonds) and from the control sample (yellow circles). }
\label{fig:compression}
\end{figure*}
However, considering the typical jump at the shock, the dawn-to-dusk electric field increases on average by a factor of 2.5 through the shock. The large drop in Dst found within 9 hours of the shock passage for 30\% of the shocks is also consistent with the compression due to the shock (and possibly the following CME or fast solar wind stream) to contribute significantly to the additional geo-effect. In short, for some events, such as the one shown in Figure~1, there are geomagnetic effects clearly caused by the shock passage, and they can be very intense.

\subsection{Shock Properties}

The left panel of Figure~\ref{fig:shock} shows the yearly distribution of the shocks as compared to that of CMEs. The number of shocks itself is known to have a solar cycle dependence \citep[]{Oh:2007}. This effect is probably amplified for shocks inside CMEs since the number of CMEs also has a solar cycle dependence. Nonetheless, we found 2 occurrences before mid-1998 and 8 after mid-2004, which are time periods of moderate activity when the monthly sunspot number was less than 50. 

The right panel Figure~5 shows the distribution of Alfv{\'e}n Mach numbers and shock speeds for the shock inside CMEs and for those from he control sample. Typical error for the Mach number and the compression ratio is $\pm 0.3$, it is $\pm 70$~km\,s$^{-1}$ for the shock speed. From this Figure, it is clear that many of the shocks propagating through CMEs have a very low Alfv\'en Mach number (45\% have $M_A < 1.75$ compared to 7\% for our control sample). However,  very few of them have slow speeds (only 1 shock propagating through a CME has a speed of less than 450~km\,s$^{-1}$, as compared to 9 from the control sample). 
The average (resp. median) speed of the shocks propagating through CME is 617 (resp. 580) km\,s$^{-1}$, significantly larger than that from the control sample (average speed of 540~km\,s$^{-1}$), and much larger than that from the other studies cited above. The average (resp. median) Alfv\'en Mach number is 2.3 (resp. 1.9), which is slightly lower than that from other samples (average of 2.9 for the control sample). In summary, these 49 shocks propagating inside CMEs are on average fast but weak, which is what was expected. This comes from the following facts: (i) only fast shocks can overtake a previous CME and propagate through it, and (ii) as the upstream and Alfv\'en speeds are high inside the CME, these shocks are weak. As noted in RC2010, what is referred to as the shock properties (shock angle, Mach numbers) is, in fact, strongly dependent on the upstream conditions.

Figure~\ref{fig:compression} shows the compression ratio with respect to the Alfv\'en Mach number and the compression ratio with respect to $\beta$. The diamonds show the 49 shocks propagating through CMEs, whereas the circles show the 45 shocks from the control sample. It appears that, at low Aflv\'en Mach numbers, shocks propagating inside CMEs have a slightly larger compression ratio as compared to shocks from the control sample at the same Mach number. We come back to these considerations in the section \ref{sec:analytical} to show how the compression ratio of a shock increases with decreasing upstream $\beta$. 
 
\subsection{Shock Location}

\begin{figure*}[t]
\centering
{\includegraphics*[width=6.2cm]{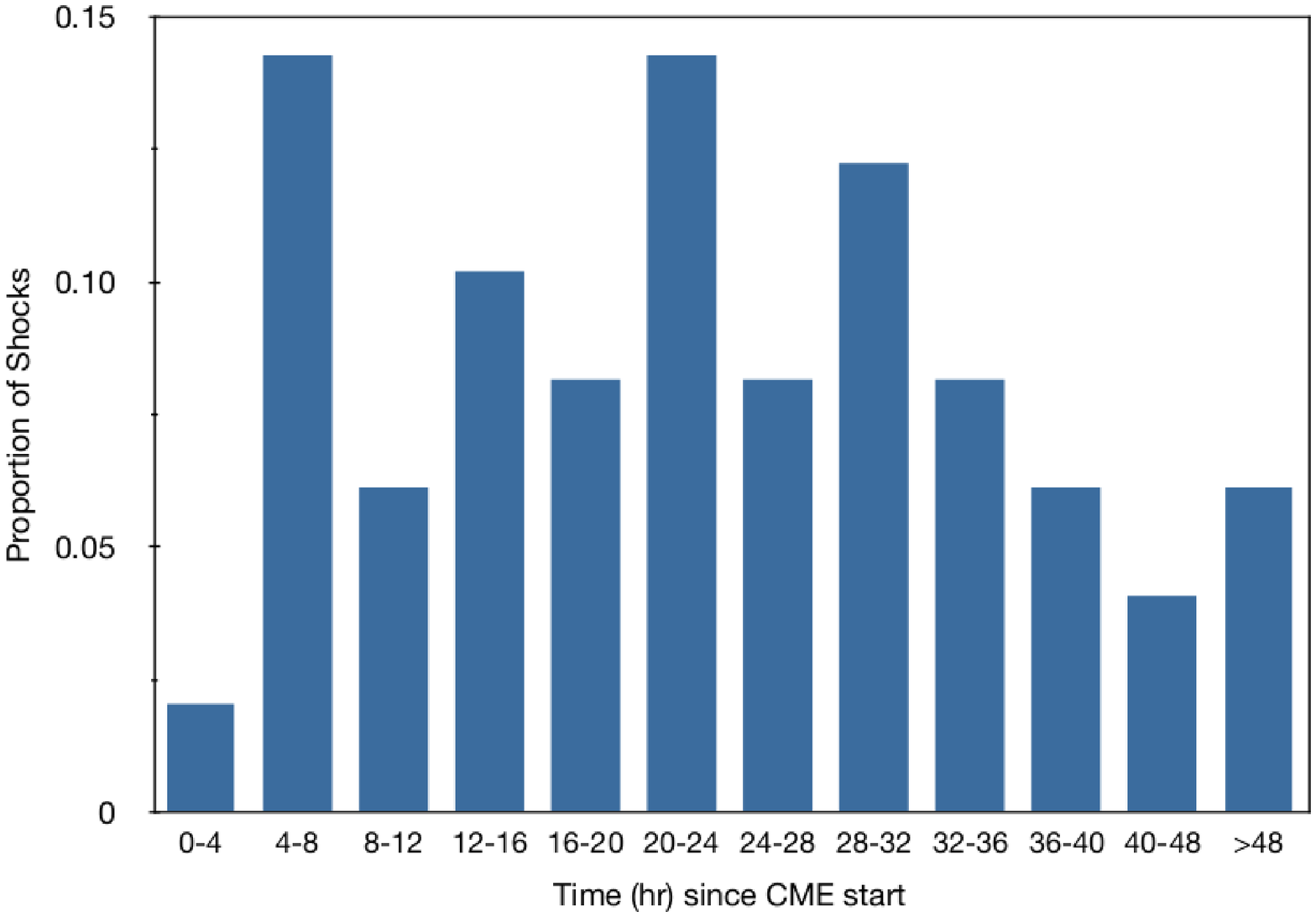}}
{\includegraphics*[width=6.4cm]{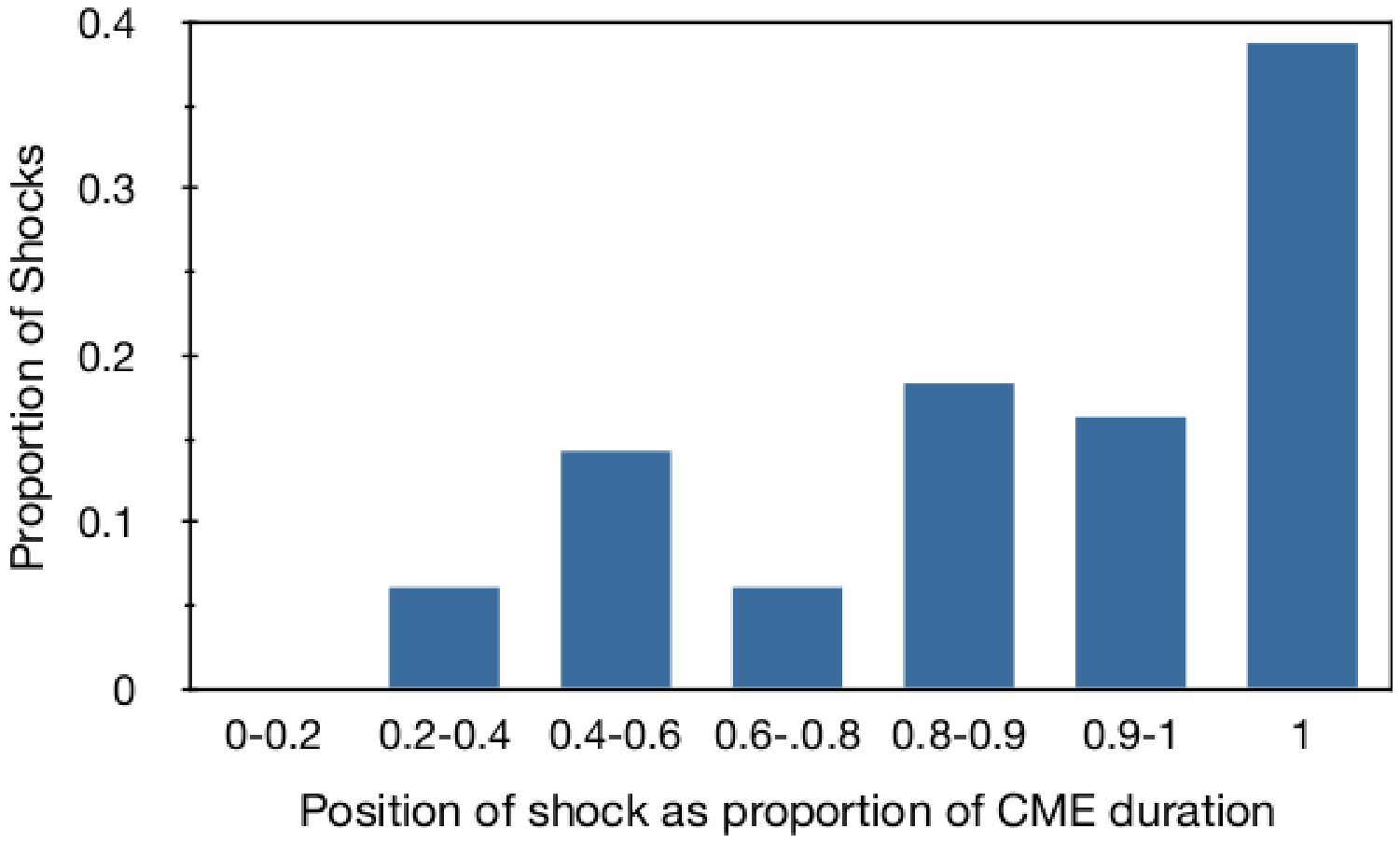}}\\
\caption{Distribution of the shock position inside the CME. In the left panel, the delay between the start of the CME and the shock location is plotted in hours; the right panel shows the location of the shock as a proportion of the CME total duration, where 1 indicates a shock at or after the back boundary.}
\label{fig:position}
\end{figure*}

In the list of CMEs of \citet{Richardson:2010b}, the median duration of a CME is about 30 hours. As shown in the left panel of Figure~\ref{fig:position}, only about one third of the shocks occur within 16 hours of the start of the CME. In fact, one third occurs more than 32 hours after the CME start. We also found no case for which the shock propagates within 3 hours of the start of the CME. 
The extreme cases correspond to very large CMEs, as for example the April 17--19, 2002 CME (duration 47 hours, shock after 40.5 hours) and the October 3--5, 2000 (duration 41 hours, shock at the back boundary). The three most extreme cases probably correspond to shocks propagating through complex ejecta composed of multiple CMEs:  the  August 20--23, 1998 CME (duration of 60 hours with a shock after 48 hours and a shock at the back boundary) and October 22-24, 2003 CME (duration of 61 hours with a shock at the back boundary); both have a radial size of more than 0.7~AU or twice more than a typical CME. 

The right panel of Figure~\ref{fig:position} shows the location of the shock within the CME as a proportion of the CME duration, where 0 identifies a shock at the CME front boundary and 1 a shock at (or after) the CME back boundary. 38\% (19/49) of the shocks occur at the back boundary of the CME. Even removing these, the distribution is clearly biased towards shocks being close to the back of the CME: the median (average) position of the 30 other shocks is 80\% (72\%) from the CME front with the closest shock 25\% of the way from the CME front (Nov. 8, 1998, see Figure~2). Only 7 shocks are in the front half of the CME, and 11 in the front 70\%. It should be kept in mind that the part of the CME downstream of the shock has been compressed, as shown for example in \citet{Lugaz:2013b}. Therefore, a shock occurring temporally halfway through the CME has already propagated through more than half of that CME. However, considering that the compression ratio of the shock is typically low, and that so few shocks are found even in the front 70\% of the CME, we believe that the distribution is truly biased.

\begin{figure}[b]
\centering
{\includegraphics*[width=9cm]{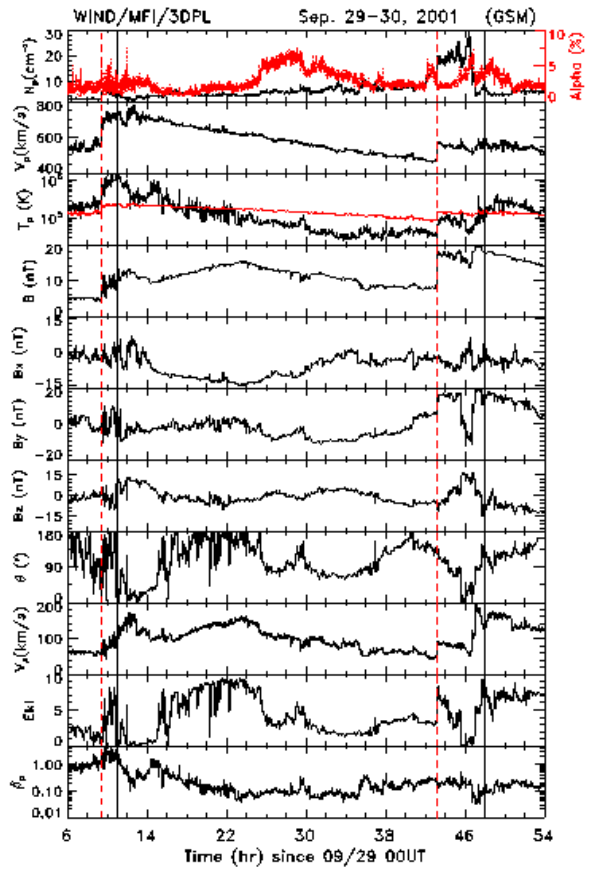}}\\
\caption{Wind observations of the shock inside a CME on 09/30/2001. The CME starts around 11:00UT on 09/29 and is preceded by a shock at 09:40UT on the same day. A second shock passes at 19:14UT on 09/30 (43.2 h).}
\label{fig:Sep30}
\end{figure}

It is our opinion that this biased distribution reflects the fact that shocks weaken due to the large Alfv{\'e}nic and upstream speeds inside CMEs and only some shocks ``survive'' the entire propagation  throughout a CME. Due to CME expansion, the upstream medium is typically faster by 100--200~km\,s$^{-1}$ at the front boundary of the CME as compared to the back. 
Many of the shocks in our sample would not be fast-mode shocks if the upstream speed increased by 100~km\,s$^{-1}$, everything else being equal. Looking at Figure~3, it is very unlikely that the September 12, 2005 shock would remain a fast-mode shock in the front half of the CME where the upstream speed is greater than 1,000~km\,s$^{-1}$, i.e. greater than the shock speed. Another example is shown in Figure~\ref{fig:Sep30} with the September 30, 2001 shock. There, a shock propagates at the back of a fast and large CME, about 43 hours after the CME start and 5 hours before its end. The shock speed at 700~km\,s$^{-1}$ is lower than the speed of the CME front ($\sim$ 750~km\,s$^{-1}$). Additionally, the Alfv\'en speed also increases by 100~km\,s$^{-1}$ to 150~km\,s$^{-1}$ from the back to the center of the CME. With a shock speed of 700~km\,s$^{-1}$, it is clear that the shock will not remain a fast-mode shock as it reaches the center of the CME. An example of a fast-mode wave, which probably is the result of a fast-mode shock decaying as it propagates through a CME was given in \citet{Wang:2003c} for the October 03--05, 2001 CME (their Figure~9). Counter-examples are the shocks shown in Figures 1 and 2 , which are likely to remain fast-mode shocks throughout the CME. 

We should also note that there are some measurements of shocks propagating through a CME sheath (for example the March 30, 2001 sequence of two shocks), which may indicate that some shocks can indeed propagate through an entire magnetic ejecta. Overall, our sample suggests that not all shocks can remain fast-mode shocks as they propagate through a CME. 
A rough estimate of the number of ``missing shocks'' can be obtained as follows. First, we cannot assume that shocks are as likely to be measured in all parts of CMEs. This is simply because, when a shock is measured inside a CME, the back portion of this CME (downstream region of the shock) has gone through compression, which shortens its duration. Consequently, it changes the probability of detecting a shock at a given location as measured from the CME total duration. For example, the shock on May 29, 2005 is measured half-way through a CME 6 hours after its start and 6 hours before its end. However, the 6 hours behind the shock, correspond to compressed CME material, not the ``actual CME'' before compression. This means that the shock has already propagated through more than half of the CME towards the front. 

The typical compression ratio, in density and magnetic field, of the shocks in our study, is 2. Therefore, we assume that the duration of the compressed portion of these CMEs has been reduced by a factor of 2 as compared to the duration before the shock passage. Then, under this assumption, we can estimate that an equal number of shocks should be measured in the front 67\% of the CMEs as in the back 33\% (based on the CME duration). 11 shocks were found in the front 67\%, 19 shocks at the CME back boundary and 19 shocks in the back 33\% (without including the shocks at the back boundary). We estimate that 25$\pm$10 shocks should have been measured in the front 67\% of the CME for the same number of shocks measured at the back of the CME. We can therefore give a rough estimate of 50\% $\pm$ 25\% of shocks which do not remain fast-mode shocks during propagation through an entire CME.

\section{Analytical Considerations}\label{sec:analytical}
In this section, we discuss the compression of a MHD shock under different upstream conditions using the analytical Rankine-Hugoniot formalism. Most of this work will be well known to researchers working on Earth's bow shock but these equations have been rarely used to discuss CME-driven shocks. 
For the sake of simplicity, we limit ourselves in the next section to shocks for which the normal is anti-parallel to the velocity vector. As CME-driven shocks tend to be relatively blunt \citep[see for example][]{Janvier:2014}, most shocks are observed with a normal within 30$^\circ$ of the GSE $x$ direction, which is, in general, the direction of the speed. This assumption is therefore justified for these simple analytical considerations.

\begin{figure}[t]
\centering
{\includegraphics*[width=7cm]{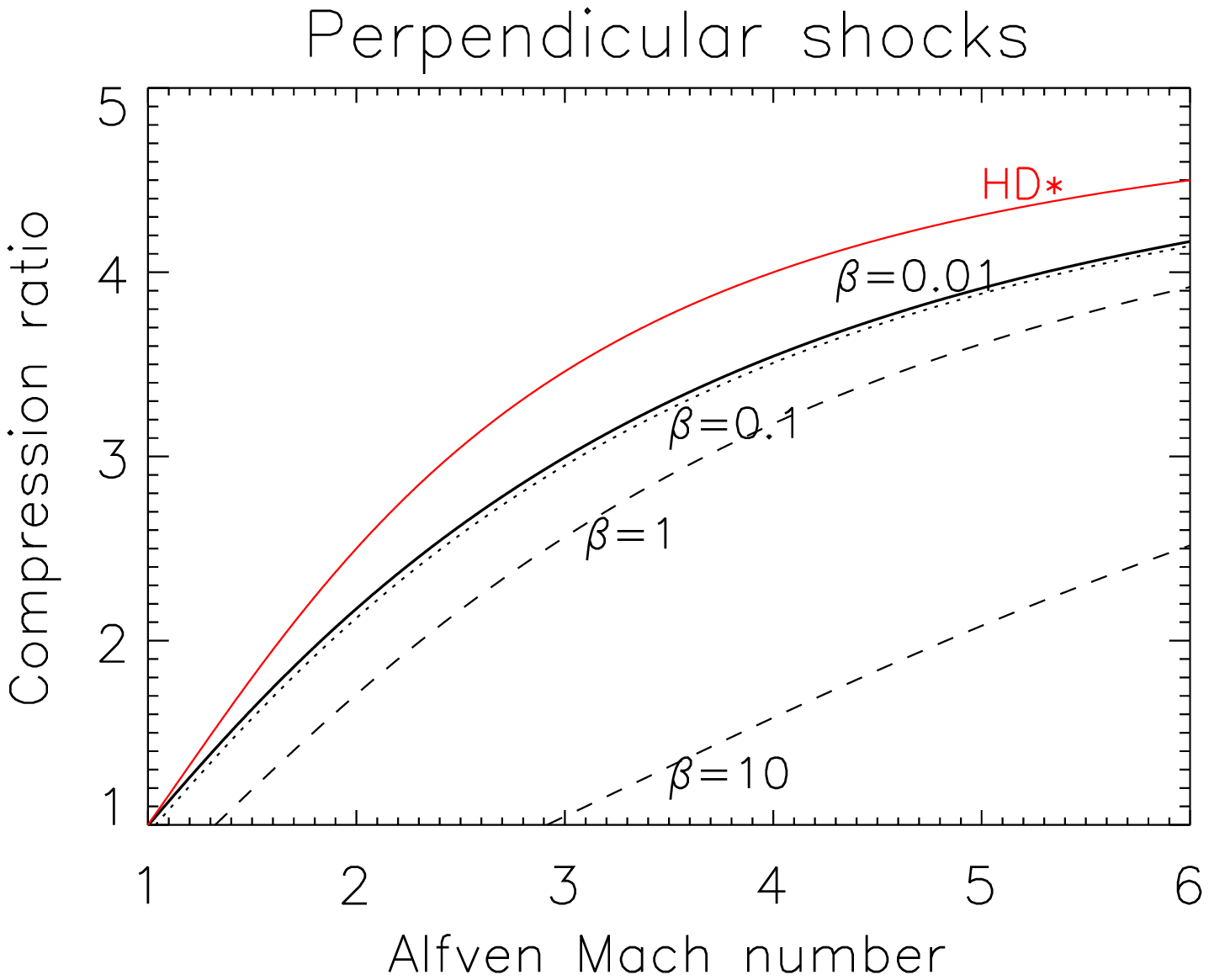}}\\
{\includegraphics*[width=7cm]{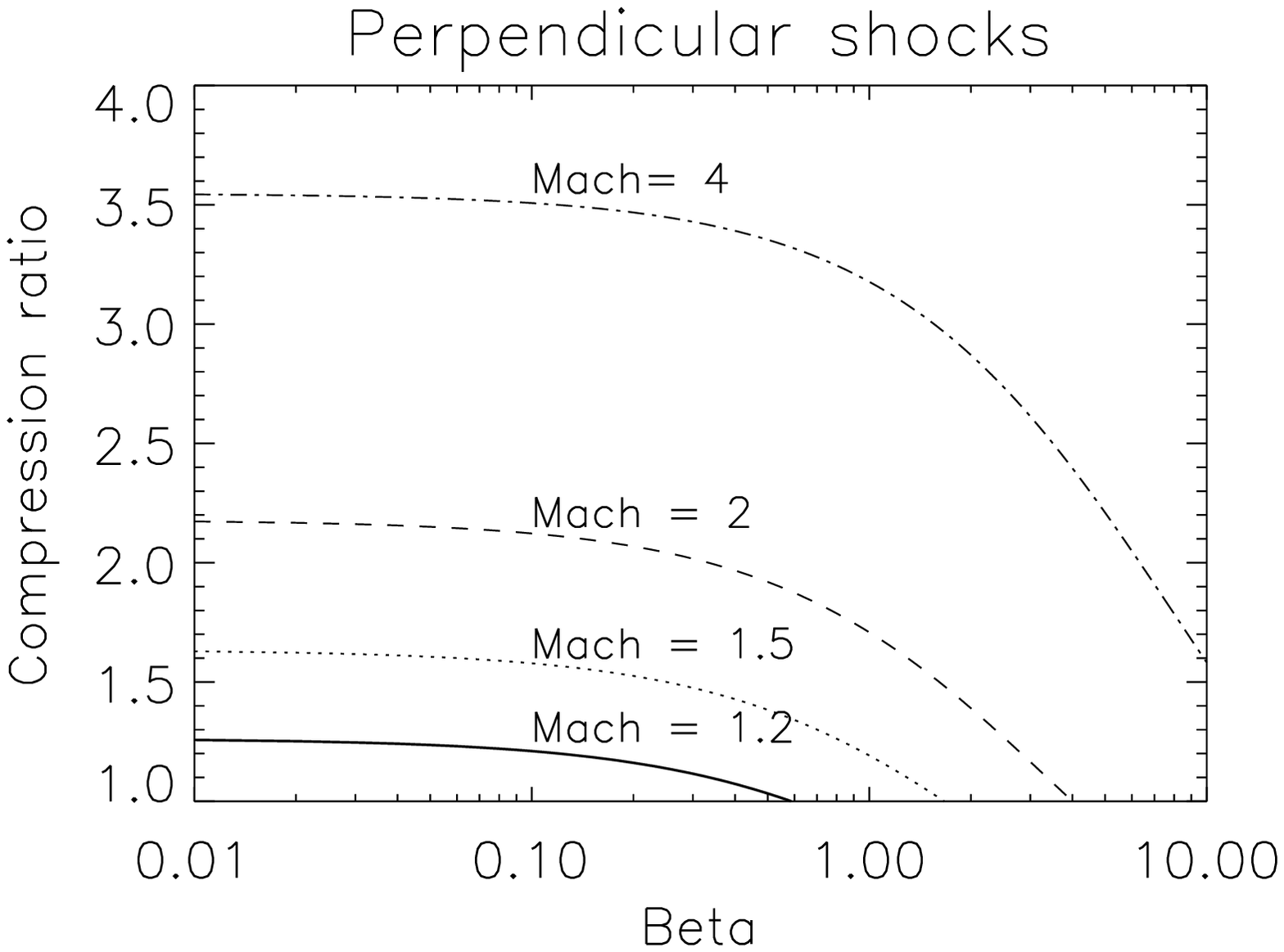}}\\
{\includegraphics*[width=7cm]{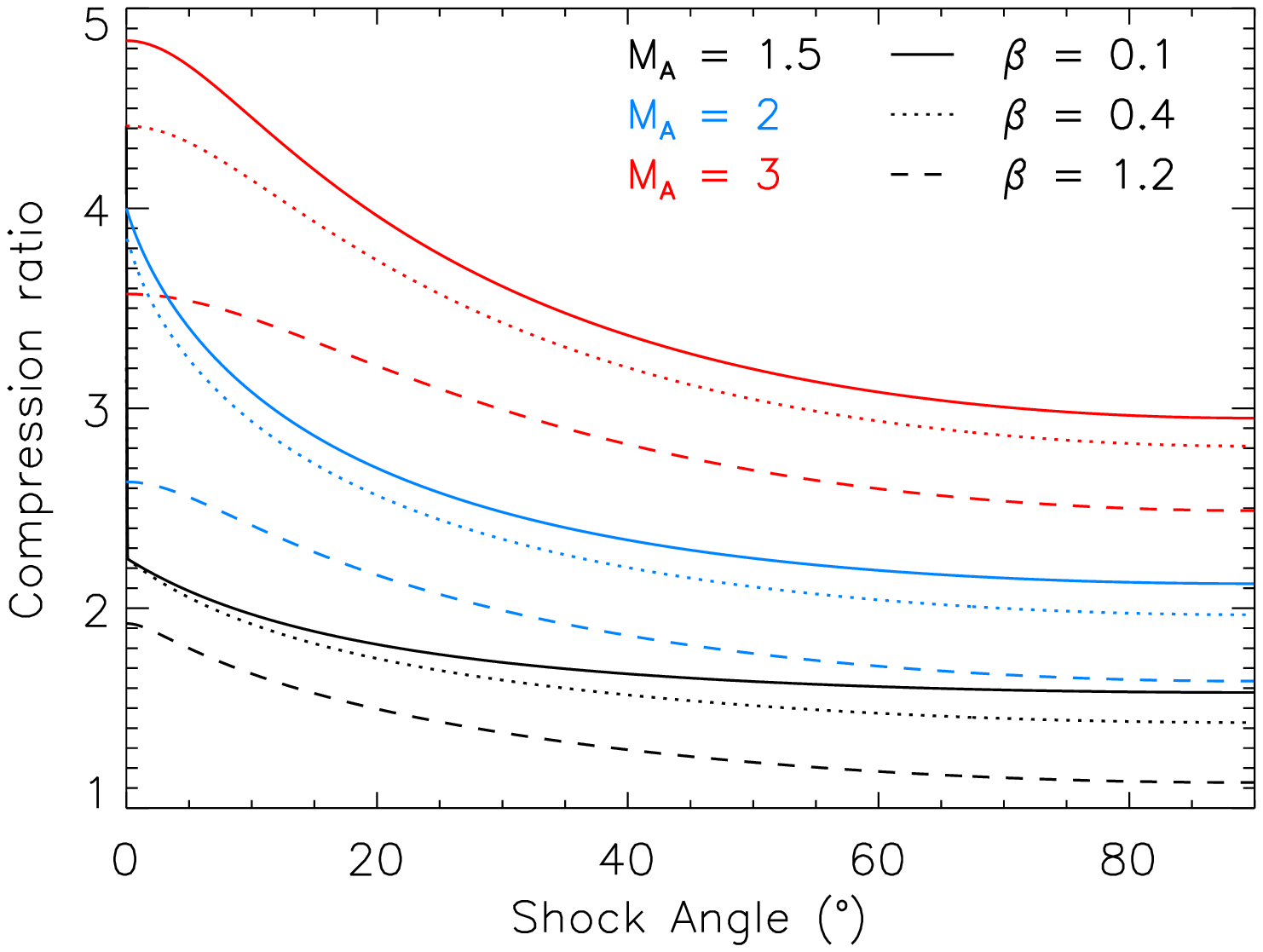}}
\caption{Compression ratio (in density) of a fast-mode shock assuming the shock normal is anti-parallel to the velocity. The first two panels show the cases of a perpendicular shock, while the last panel shows the dependence of the compression ratio on the shock angle, Alfv{\'e}n Mach number and plasma $\beta$.}
\label{fig:comp}
\end{figure}

It is important to remember that the plasma $\beta$ is related to the ratio of the Alfv{\'e}n, $M_A$ to sonic, $M_s$ Mach numbers as follows:
\begin{equation}
\frac{M_A^2}{M_s^2} = \frac{\gamma \beta}{2}
\end{equation}
Under this assumption, the formula for the compression ratio, $1/X$ of a MHD shock is the positive root of the cubic equation \citep[]{Zhuang:1981,Cairns:1994}: 

\begin{equation}
aX^3 - b X^2 + c X - d = 0, 
\end{equation}
with:
\begin{eqnarray*}
a & = & (\gamma + 1) M_A^6 \\
b & = & (\gamma -1) M_A^6 + (\gamma + 2) M_A^4 \cos^2 \theta + \gamma (1 + \beta) M_A^4 \\
c & = & (\gamma - 2 + \gamma \cos^2 \theta) M_A^4 + (\gamma + 2 \gamma \beta  + 1) M_A^2 \cos^2 \theta\\
d & = & (\gamma - 1) M_A^2 \cos^2 \theta + \gamma \beta \cos^4 \theta
\end{eqnarray*}
where $M_A$ is the Alfv{\'e}n Mach number and $\theta = \theta_\mathrm{Bn} $ the angle between the shock normal and the upstream magnetic field vector. The full version of this equation is a more complex cubic equation as discussed in \citet{Petrinec:1997} and \citet{Kabin:2001}, which involves the angle between the shock normal and the velocity vector.

To start the discussion, we can make a further simplification, assuming a perpendicular shock ($\theta_\mathrm{Bn} = 90^\circ$). The equation simplifies to a quadratic equation, since $d = 0$. For $\gamma < 2$, the equation always has a unique positive root, $X_\perp$. However, the shock must also be compressive ($X_\perp < 1$) to be an acceptable solution to the Rankine-Hugoniot relations, which can be written as:
$$ (\gamma + 1) M_A^2 - (\gamma -1) M_A^2 - \gamma (1 + \beta)+ \gamma - 2 > 0$$ or
\begin{equation}
M_A^2 > 1 + \frac{\gamma \beta}{2} \label{eq:limit_perp}
\end{equation}
Equation \ref{eq:limit_perp} clearly shows that MHD shocks with extremely low Alfv{\'e}n Mach numbers can only be observed when the plasma $\beta$ is low. For example, for $\gamma = 1.5$, the minimum Alfv{\'e}n Mach number at $\beta = 1$ is 1.32; whereas for $\beta = 0.1$, it is 1.04. Below this limit, the shock fast magnetosonic Mach number becomes less than 1. Talking only about the fast magnetosonic Mach number or the Alfv{\'e}n Mach number hides the fact that some of these shocks have large sonic Mach numbers. 
The left and middle panels of Figure~\ref{fig:comp} show the compression ratio of a perpendicular shock for different values of $\beta$ and the Alfv{\'e}n Mach number, illustrating how weak shocks may be significantly more compressive for $\beta \le 0.1$.

Examples of the solutions to equation~(2) for oblique shocks are shown in the last panel of Figure~\ref{fig:comp}, where the compression ratio is shown for different values of $\beta = [0.1, 0.4, 1.2]$ and for $M_A = [1.5, 2, 3]$. For $\beta = 0.01$, the values are nearly identical to that for $\beta = 0.1$. Note also that the sonic Mach number is the same for the solid black curve ($\beta = 0.1, M_A = 1.5$) and the dotted red curve ($\beta = 0.4, M_A = 3$), and the fast  Mach number is about the same between the solid black curve and the blue dashed curve ($\beta = 1.2, M_A = 2$). 
In a low-$\beta$ plasma, a shock with an Alfv{\'e}n Mach of 1.5 has a compression ratio between 1.6 and 2, depending on the shock angle. For $\beta = 1$, the compression ratio is between 1.1 and 1.6. This is somewhat consistent with the behaviors found for the 49 shocks inside CMEs as compared to the 45 shocks in more typical upstream conditions, as shown in the left panel of Figure~6. 

The compression of the tangential component of the magnetic field is given by $1/X_\mathrm{Bt}$ with $X_\mathrm{Bt} = X \frac{M_A^2 - X\cos^2\theta}{M_A^2 - \cos^2 \theta}$. It is clear that the tangential magnetic compression is always larger than the compression in density, the two being equal for perpendicular shocks. For small Alfv{\'e}n Mach and quasi-parallel situation, the solution tends towards the switch-on shock. It should be remembered that for shocks inside CMEs, the shock angle depends primarily on the local orientation of the magnetic field. It is likely that the shock angle dramatically changes as the shock propagates inside the CME. 

\section{Discussions and Conclusion} \label{sec:conclusion}

In this article, we identified 49 shocks propagating inside or closely following a CME during solar cycle 23 (1997--2006, with no such shock during 2007--2008). We analyzed the properties of these shocks and compared them to 45 shocks propagating into normal solar wind conditions. We found that shocks propagating inside CMEs are found to be, as expected, fast and weak with a median speed of 580~km\,s$^{-1}$ and a median Alfv{\'e}n Mach number of 1.9. These shocks represent about 15-20\% of all shocks measured at 1~AU during solar cycle 23. RC2010 reported that 30\% of quasi-perpendicular shocks during the same time period were found inside or following CMEs. About 70\% of the shocks in our sample have a shock angle between 65$^\circ$ and 90$^\circ$, i.e. are quasi-perpendicular (however, 24 out the 49 shocks analyzed here were not part of the RC2010 study). As detailed in RC2010, the fact that shocks inside CMEs represent a larger proportion of quasi-perpendicular shocks than that of the whole shock population is a direct consequence of the conditions required for a shock to be quasi-perpendicular at 1~AU. If  a spacecraft crosses a shock close to its nose, quasi-perpendicular conditions can only be met if the magnetic field is primarily in the $y-z$ plane. This is often the case inside a CME but rarely outside of one. When the ``wing'' of a shock is measured, the shock may be quasi-perpendicular under normal solar wind conditions. However, such measurements are rarer \citep[]{Janvier:2014}. In short, shocks inside CMEs have a higher likelihood of being quasi-perpendicular than random shocks. 

Compared to the full list of CMEs measured at 1~AU, we find that about 9\% of CMEs have a shock propagating through them, and an additional 6\% have a shock at the location of the back boundary \citep[following the boundaries selected by][]{Richardson:2010}. This proportion varies significantly with solar cycle. For 2003, when the total number of CMEs measured at Earth was relatively low, but many occurred during a short time period (the October-November months, the so-called Halloween storms), as many as 30\% of the CMEs identified at Earth had a shock inside them. 

Large southward magnetic fields are the leading cause of geomagnetic storms, and these occur preferentially inside CMEs. Here, we found that 19 occurrences of shocks inside CMEs were associated with an intense geomagnetic storm within 12 hours of the shock passage. This represents over 20\% of all intense geomagnetic storms identified in solar cycle 23. Shocks inside CMEs are therefore an important cause of intense geomagnetic storms at 1~AU. In four cases, the downstream magnetic field a few hours after the shock was larger than 30~nT. In addition, the August 24, 2005 CME has a sharp discontinuity 3 hours after the shock, which compressed the magnetic field from 25~nT to close to 60~nT. These large values of the magnetic field strengths are comparable to that of CMEs driving superstorms (Dst$_\mathrm{peak} < -250$~nT). However, here they are caused by weak-to-average CMEs. This shows that shocks propagating inside CMEs are one of the ways to create strong $B_z$ (and strong magnetic field) by compressing weak or average CMEs into CMEs with more extreme values of the magnetic field strength. 

We did not systematically investigate the source of the shocks propagating inside CMEs, but we believe that most of them are CME-driven shocks for the studied period. First, most shocks at 1~AU are CME-driven. For example, \citet{Jian:2006} and \citet{Jian:2006b} reported 151 CME-driven shocks from 1995 to 2004 as compared to a total of 67 forward shocks driven by SIRs or CIRs during the same time period. \citet{Oh:2007} found a lower proportion of CIR-driven shocks at about 16\% from 1995 to 2001. They also found that CIR-driven shocks at 1~AU were significantly weaker and slower than CME-driven shocks (420 as compared to 500~km\,s$^{-1}$). This makes it unlikely that CIR-driven shocks are able to overtake and propagate through CMEs, although they may contribute to the large number of shocks propagating at the back of CMEs. In addition, we find that many CMEs into which the shocks propagate are complex (for example, about a quarter of them have a duration of more than 30 hours), which may indicate that they are in fact the result of the interaction of multiple CMEs. 

We have also analyzed the location of the shocks within the CMEs. Most shocks occur more than 20 hours after the start of the CME in the back half of the magnetic ejecta. Even taking into consideration the compression of the back of the CME by the shock, it appears that fewer shocks are measured in the front part of the CME as what could be expected from an uniform distribution. We conclude that about half of the shocks entering a CME may not remain fast-mode shocks throughout the CME. This is likely due to the larger upstream solar wind speed, due to the CME expansion, and the location of the maximum in the Alfv\'enic speed, which often occurs close to the middle of the CME. In the CME frame, the shock is often near-stationary, while the back of the CME propagates sunwards towards the shock due to the CME expansion. Future work should further investigate possible signatures of shocks after they exit CMEs.

\begin{acknowledgments}
The authors would like to thank the reviewers for their comments. The authors acknowledge the use of the NASA/GSFC OMNIWEB and CDAWEB data and the shock list from CfA maintained by M. Stevens (\url{https://www.cfa.harvard.edu/shocks/}). 

The research for this manuscript was supported by the following grants: NSF AGS-1239699, NASA NNX15AB87G, NNX13AP39G and  NNX13AP52G.  N.~L. would like to thank the participants to the RAL Space CME-CME interaction workshop in March 2014 for fruitful discussion about shock propagating inside CMEs. C.W.S. is supported by Caltech subcontract 44A-1062037 to UNH in support of the ACE/MAG instrument.
\end{acknowledgments}
\newpage

 \appendix
\section{Control Sample}

\begin{figure*}[h]
\centering
{\includegraphics*[width=7.cm]{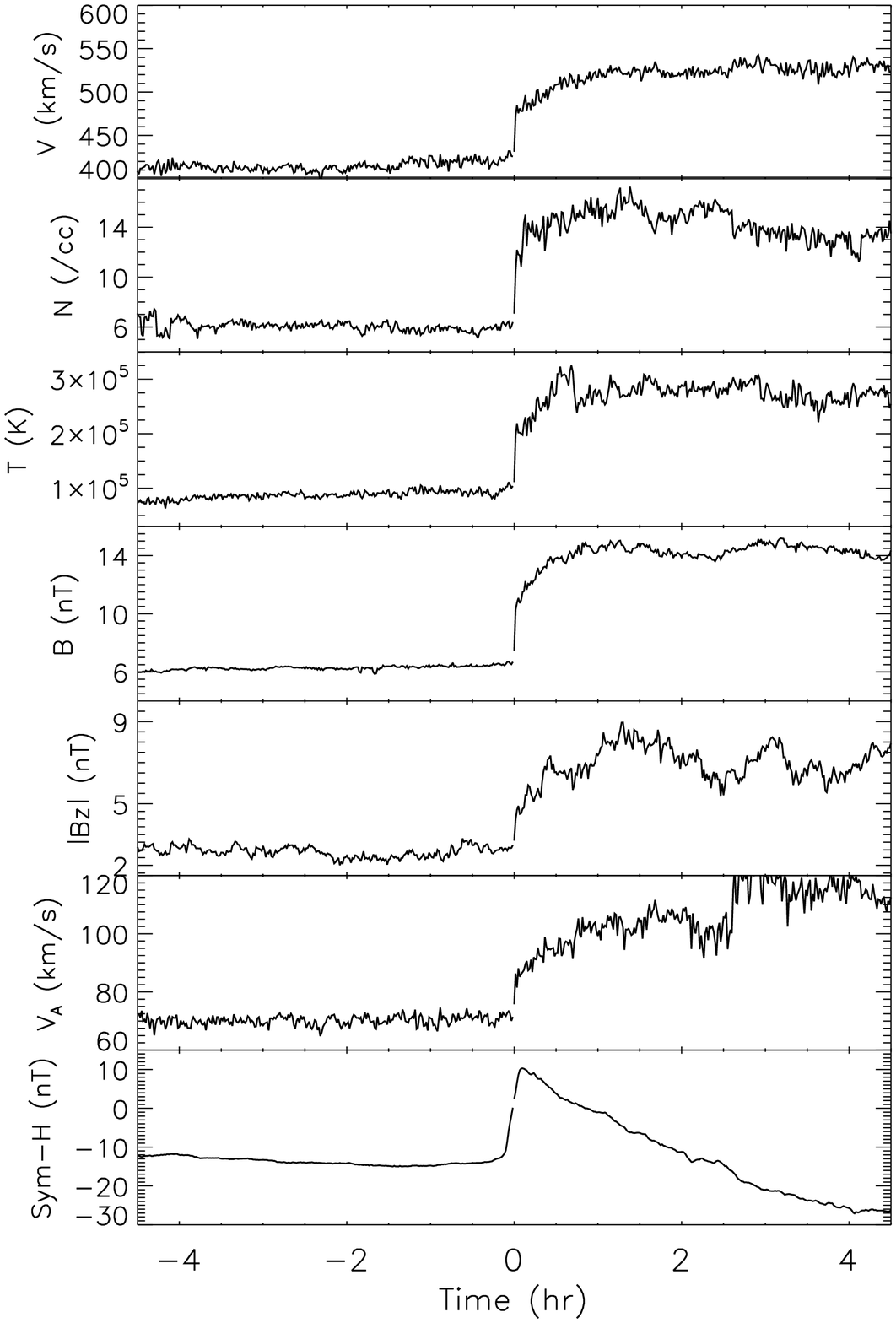}}
{\includegraphics*[width=7.cm]{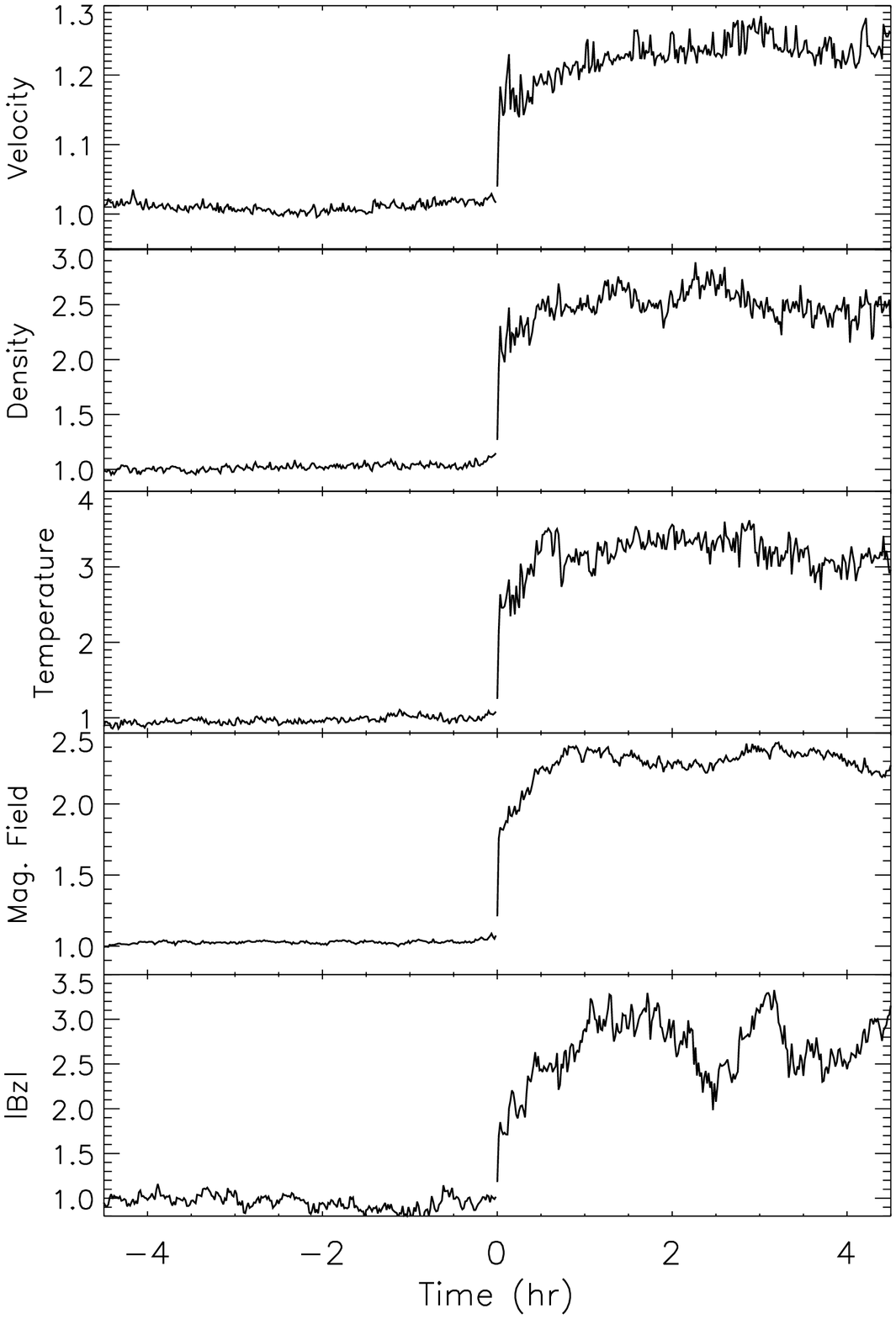}}\\
\caption{Superposed epoch analysis of the 45 shocks from our control sample. The left panels show the velocity, proton density, proton temperature, total magnetic field, absolute value of the magnetic field $B_z$ component in GSM coordinates, Alfv{\'e}n speed and Sym-H index, from top to bottom. The right panels show the dimensionless velocity, density, temperature, total and $B_z$ magnetic field, from top to bottom.}
\label{fig:superposed_control}
\end{figure*}

The yearly distribution of the 45 shocks from our control sample is shown in the left panel of Figure~5. The main properties of these shocks are shown in Figures~5 and 6. Figure~\ref{fig:superposed_control} shows the superposed epoch analysis for these 45 shocks in the same format as Figure~\ref{fig:superposed}. Table~2 lists the properties of these shocks. 

\newpage

\bibliographystyle{agu}

\end{article}

\newpage 
\begin{landscape}
\begin{longtable}{|ccc|ccccccc|cccc|ccc|}
\hline
\multicolumn{3}{|c|}{CME} & \multicolumn{7}{|c|}{Shock} & \multicolumn{4}{c|}{Upstream} & \multicolumn{3}{c|}{Geo-effect}  \\
\hline
Start & End & MC & Time & $\Delta$t & V & $M_A$ & $\Theta_{\mathrm{Bn}}$ & X$_\mathrm{N}$ & X$_\mathrm{B}$ & V$_\mathrm{fast}$ & V$_\mathrm{Alfv}$ & $\beta$ & V$_\mathrm{up}$ & Dst$_\mathrm{min}$ & $\Delta$Dst& $\Delta$ t\\
\hline
\multicolumn{17}{|c|}{\bf 1997}\\
\hline
10/10 11:00 & 10/10 22:00 &1 &10/10 15:57 & {\bf 4} & 470 & 1.6 & 85 & 1.6 & 1.6 & 55 & 50 & 0.09 & 410  &-100 &-75 & 9\\
\hline
\multicolumn{17}{|c|}{\bf 1998}\\
\hline
05/02 05:00 & 05/04 02:00 & 2 & 05/03 17:02 & {\bf 36} & 500 & 1.9 & 40 & 3.2 & 3.1 & 40 & 40 & 0.07 & 430 & -162 & -103 & 12\\
08/05 13:00 & 08/06 12:00 & 1 & 08/06 07:16 & {\bf 18.5} & 470 &1.5 &  85 & 1.8 & 1.8 & 70 & 70 & 0.07 & 370 & -138 & -107 & 5\\
11/07 22:00 & 11/09 01:00 & 1 & 11/08 04:41 & {\bf 6.5} & 640 & 1.4 & 60 & 1.9 & 1.9 & 175 & 170 & 0.03 & 450 & -149 & -40 & 2.3\\
\hline
\multicolumn{17}{|c|}{\bf 1999}\\
\hline
02/16 15:00 & 02/17 11:00 & 1 & 02/17 07:12 & {\bf 16} & 555 & 1.6 & 85 & 1.5 & 1.4 & 70 & 60 & 0.2 & 465 & -32 & -27 & 5 \\
02/17 16:00 & 02/18 10:00 & 0 & 02/18 02:48 & {\bf 11} & 680 & 3.4 & 50 & 2.9 & 2.5 & 100 & 85 & 0.3 & 390 & -123 & -97 & 8 \\
07/07 07:00 & 07/08 04:00 & 0  & 07/08 03:47 & 21 & \_ & \_ & 80 & \_ & 1.6 & 95 & 90 & 0.07 & 420 & -10 & -19 & 12 \\
08/20 23:00 & 08/23 11:00 & 1 & 08/22 23:27 & {\bf 48.5} & 455 & 1.3 & 65 & 1.3 & 1.3 & 105 & 100 & 0.15 & 410 & -66 & -21 & 3 \\
08/20 23:00 & 08/23 11:00 & 1 & 08/23 12:11 & 61 & 470 & 1.4 & 60 & 1.6 & 1.6 & 70 & 70 & 0.04 & 385 & -52 & 11 & 6 \\
11/12 10:00 & 11/13 18:00 & 0 & 11/13 12:48 & {\bf 27} & 465 & 1.4 & 85 & 1.8 & 1.8 & 85 & 80 & 0.05 & 410 & -101 & -48 & 9 \\
\hline
\multicolumn{17}{|c|}{\bf 2000}\\
\hline
07/11 02:00 & 07/11 14:00 &  0 & 07/11 12:15 & {\bf 10} & 550 & 2.2 & 90 & 1.9 &1.9 & 85 & 80 & 0.1 & 430 & -21 & -48 & 11 \\
07/14 17:00 & 07/15 14:00 & 2 & 07/15 14:35 & 21.5 & 1200 & 7.5 & 25 & 4 & 2.3 & 100 & 100 & 0.1 & 650 & -289 &  -234 & 8 \\
07/27 02:00 & 07/28 02:00 & 1 & 07/28 06:38 & 28.5 & 460 & 1.7 & 50 & 2.6 & 2.6 & 80 & 80 & 0.05 & 340 & -46 & -12 & 12\\
08/10 19:00 & 08/11 21:00 & 2 & 08/11 18:49 & {\bf 24} & 580 & 1.6 & 70 & 2.2 & 2.0 & 175 & 170 & 0.03 & 410 & -72 & -10 & 12 \\
10/03 10:00 & 10/05 03:00 & 2 & 10/05 03:28 & 41.5 & 510 &2.9 & 70 & 2.3 & 2.2 & 65 & 55 & 0.1 & 360 & -175 & -53 & 5.5\\
11/27 08:00 & 11/28 03:00 & 0 & 11/28 05:25 & 21.5 & 580 & 2.2 & 55 & 2.3 & 2.2 & 75 & 65 & 0.25 & 525 & -73 & -27 & 4\\
11/28 11:00 & 11/29 22:00 & 1 & 11/29 05:47 & {\bf 18.5} &  575 &1.3 &  75 & 1.4 & 1.4 & 75 & 70 & 0.15 & 540 & -109 & -24 & 8\\
\hline
\multicolumn{17}{|c|}{\bf 2001}\\
\hline
04/11 22:00 & 04/13 07:00 & 2 & 04/13 07:23 & 37.5 & 750 & 1.5 & 160 & 1.3 & 1.3 & 110 & 110 & 0.05 & 580 & -77 & -11 & 9\\
08/30 17:00 & 08/31 10:00 & 1 & 08/31 01:25 & {\bf 8.5} & 475 &1.5 & 80 & 1.4 & 1.3 & 75 & 70 & 0.3 & 400 & -40 & -32 & 9 \\
09/13 18:00 & 09/14 22:00 & 1 & 09/14 01:59 & {\bf 8} & 490 & 2.8 & 65 & 2.6 & 2.5 & 45 & 40 & 0.2 & 370 & -3 & 1 & 12 \\
09/25 06:00 & 09/25 20:00 & 0 & 09/25 20:17 & 14.5 & 855 & 5 & 65 & 3 & 2.5 & 90 & 85 & 0.05 & 370 & -102 & -89 & 6 \\
09/29 11:00 & 10/01 00:00 & 1& 09/30 19:14 & {\bf 32} & 700 & 5 & 60 & 1.8 & 1.7 & 60 & 50 & 0.3 & 460 & -112 & -57 & 12 \\
10/27 03:00 & 10/28 12:00 & 0 & 10/28 03:13 & {\bf 24} & 590 & 2.3 & 60 & 2.7 & 2.3 & 110 & 105 & 0.04 & 355 & -157 & -140 & 9 \\
10/29 22:00 & 10/31 13:00 & 0 & 10/31 13:47 & 40 & 440 & 2.5 & 35 & 2.1 & 1.7 & 50 & 50 & 0.05 & 325 & -81 & -62 & 11\\
11/05 19:00 & 11/06 06:00 & 1 & 11/06 01:25 & {\bf 6.5} & \_& \_ & 80 & \_ &  3 & 140 & 135 & 0.03 & 420 & -292 & -219 & 5.5\\
12/28 00:00 & 12/29 12:00 & 0 & 12/29 05:17 & {\bf 29.5} & 530 & 2.9 & 40 & 3.5 & 2.9 &55 & 50 & 0.1 & 375 & -23 & -35 & 12 \\
12/30 00:00 & 12/30 18:00 & 1 & 12/30 20:05 & 20 & 530 & 2.2 & 55 & 2.5 & 2.5 & 135 & 120 & 0.4 & 425 & -46 & -31 & 12\\
\hline
\multicolumn{17}{|c|}{\bf 2002}\\
\hline
03/19 05:00 & 03/20 16:00 & 2 & 03/20 13:20 & {\bf 32.5} & 730  & 1.2 & 130 & 2.9 & 1.2 & 420 & 420 & 0.005 & 400 & 4 & 22 & 12 \\
03/21 14:00 & 03/22 06:00 & 0 & 03/22 03:52 & {\bf 14} & 640 & 1.4  & 110 & 1.5 & 1.5 & 70 & 65 & 0.07 & 430 & -3 & -7 & 12\\
04/17 16:00 & 04/19 15:00 & 2 & 04/19 08:25 & {\bf 40.5} & 780 & 1.6  & 70 & 2.5 & 2.5 & 200 & 190 & 0.05 & 465 & -98 & -36 & 6.5\\
07/18 12:00 & 07/19 09:00 & 1 & 07/19 10:09 & 22 & 510 & 1.1 & 90 & 2.3 & 2.2 & 95 & 85 & 0.3 & 430 & -10 & -8 & 4 \\
08/01 09:00 & 08/01 23:00 & 2 & 08/01 23:09 & 14 & 495 & 1.6 & 60 & 1.6 & 1.5 & 70 & 65 & 0.09 & 435 & -102 & -85 & 8 \\
09/07 12:00 & 09/08 04:00 & 1 & 09/07 16:22 & {\bf 4.5} & 630 & 2.4 & 90 & 2.9 & 2.9 & 105 & 100 & 0.03 & 400 & -142 & -97 & 4\\
\hline
\multicolumn{17}{|c|}{\bf 2003}\\
\hline
05/29 13:00 & 05/29 23:00 & 1 & 05/29 18:31 & {\bf 5.5} & 850 & 2 & 90 & 1.9 & 1.9 & 130 & 120 & 0.08 & 670 & -144 & -90 & 5.5 \\
05/30 02:00 & 05/30 16:00 & 1 & 05/30 15:53 & 14 & 850 & 1.5 & 110 & 2.0 & 1.9 & 190 & 185 & 0.05 & 575 & -54 & 25 & 8\\
08/16 02:00 & 08/17 16:00 & 1 & 08/17 13:41 & {\bf 33.5} & 600 & 5 & 90 & 2.1 & 2.1 & 65 & 60 & 0.04 & 420 & -63 & -77 & 10 \\
10/22 02:00 & 10/24 15:00 & 1 & 10/24 15:45 & 62 &  650 & 3 & 60 & 2.3 & 2.1 & 85 & 80 & 0.2 & 450 & -36 & 8 & 9 \\
10/25 14:00 & 10/26 04:00 & 0 & 10/26 08:55 & 19 & \_ & \_ & \_ & \_ & 1.2  &225 & 220 & 0.03 & 380 & -8 & 16 & 12\\
10/25 14:00 & 10/26 04:00 & 0 & 10/26 19:28 & 29.5 & 640 & 1.4 & 95 & 1.4 & 1.4 & 185 & 180 & 0.07 & 450 & -52 & -47 & 10 \\
10/26 22:00 & 10/28 00:00 & 0 & 10/28 02:25 & 28.5 & \_ & \_ & 70 & \_ & 1.5& 200 & 190 & 0.15 & 460 & -32 & -8 & 8 \\
\hline
\multicolumn{17}{|c|}{\bf 2004}\\
\hline
01/22 08:00 & 01/23 17:00 & 0 & 01/23 15:00 & {\bf 31} &\_ & \_ &\_ & 1.3  & 1.0 & 115 & 110 & 0.1 & 480 & -78 & -31 & 3.5 \\
07/25 20:00 & 07/26 22:00 & 1& 07/26 22:25 & 26.5 & 1100 & 8 & 55 & 5 & 3.5 & 90 & 60 & 2 & 605 & -138 & -96 & 8.5 \\
11/07 22:00 & 11/09 10:00 & 2 & 11/09 09:25 & 35.5 & 800 & 4 & 45 & 3.9 & 2.3 & 110 & 105 & 0.08 & 580 & -139 & -35 & 8\\
\hline
\multicolumn{17}{|c|}{\bf 2005}\\
\hline
05/29 03:00 & 05/29 15:00 & 1 & 05/29 08:59 & {\bf 6} & 630  & 2 & 50 & 2.4 & 2 & 90 & 85 & 0.04 & 370 & -3 & 29 & 11 \\
06/15 05:00 & 06/16 09:00 & 2 & 06/16 08:09 & 27 & 620 & 2.2 & 70 & 1.5 & 1.5 & 115 & 110 & 0.02 & 465 & -30 & -39 & 9\\
08/24 00:00 & 08/24 11:00 & 1 & 08/24 05:35 & {\bf 5.5} & 570 &2 &  85 & 2.5 & 2.4 & 80 & 70 & 0.3 & 460 & -184 & -197 & 6 \\
09/11 05:00 & 09/12 07:00 & 0 & 09/12 06:00 & 25 & 1040 & 1.6 & 20 & 3 & 1.3 & 190 & 190 & 0.04 & 740 & -77 & -9 & 5 \\
\hline
\multicolumn{17}{|c|}{\bf 2006}\\
\hline
12/31 04:00 & 01/01 17:00 & 2 & 01/01 13:26 & {\bf 33.5} & 450 &1.6 &  90 & 1.6 & 1.6 & 90 & 90 & 0.08 & 440 & 9 & 4 & 11\\
12/15 20:00 & 12/16 19:00 & 0 & 12/16 17:34 & 21.5 & 700 &1.6 &  75 & 2.3 & 2.3 & 70 & 65 & 0.15 & 560 & -36 & 15 & 12\\
\hline
\multicolumn{3}{|c|}{\bf Average} & & 23.8  & 630 & 2.4 & 70 & 2.2 & 1.95  & 112 & 107 & 0.15 & 455 & $-82$  & $-46$ & \\
\multicolumn{3}{|c|}{\bf Median} & & 22 & 585& 1.8 & 70 & 2.1 & 1.9 & 90 & 85 & 0.08 & 430 & $-72$ & $-32$ & \\
\hline
\caption{List of 49 shocks propagating inside CMEs from 1997 to 2006. The columns list the CME start and end times, and type as given by \citet{Richardson:2010} (2 = MC, 1 = MC-like, 0 = irregular CME); the shock time, the time delay (in hours) since the start of the CME (bolded $\Delta\,t$ indicates that the shock time occurred before the CME end time), the shock speed, Alfv{\'e}n Mach number, angle, compression ratios in density and magnetic field; the upstream fast magnetosonic, Alfv{\'e}n speed, $\beta$ and solar wind speed; the minimum Dst and drop in Dst in the 12 hours following the shock passage and the delay in hour  between the shock arrival and the minimum Dst.}
\end{longtable}
\end{landscape}

\newpage

\begin{longtable}{|cccccc|ccc|}
\hline
\multicolumn{6}{|c|}{Shock} & \multicolumn{3}{c|}{Upstream}  \\
\hline
Time & V & $M_A$ & $\Theta_{\mathrm{Bn}}$ & X$_\mathrm{N}$ & X$_\mathrm{B}$ & V$_\mathrm{Alfv}$ & $\beta$ & V$_\mathrm{up}$ \\
\hline
\multicolumn{9}{|c|}{\bf 1997}\\
\hline
09/02 22:54 & 370 & 3 & 50 & 2 & 1.9 & 25 & 0.7 & 315\\
\hline
\multicolumn{9}{|c|}{\bf 1998}\\
\hline
04/30 09:23 & 370 & 4 & 165 & 4.5 & 1.5 & 7 & 3.8 & 320\\
08/10 00:46 & 480 & 2 & 60 & 1.9 & 1.8 & 55 & 0.15 & 395\\
10/23 13:12 & 580 & 2 & 45 & 2.3 & 2.3 & 55 & 0.7 & 490\\
\hline
\multicolumn{9}{|c|}{\bf 1999}\\
\hline
01/13 10:55 & 430 & 2.3 & 85 & 2.1 & 2.1 & 35 & 0.5 & 345\\
01/22 20:27 & 650 & 1.6 & 30 & 1.5 & 1.5 & 125 & 0.2 & 515\\
07/02 01:04 & 560 & 1.8 & 70 & 2 & 2 & 85 & 0.3 & 465\\
09/12 03:58 & 520 & 3 & 70 & 2.5 & 2.3 & 40 & 0.5 & 420\\
09/15 08:00 & 650 & 2.1 & 70 & 2.1 & 2 & 80 & 0.65 & 515\\
10/21 02:23 & 470 & 2.4 & 75 & 2.5 & 2.5 & 55 & 0.15 & 350 \\
\hline
\multicolumn{9}{|c|}{\bf 2000}\\
\hline
02/14 07:36 & 680 & 1.6 & 65 & 1.8 & 1.8 & 75 & 0.45 & 570\\
04/06 16:45 & 650 & 4 & 70 & 3.7 & 3.6 & 65 & 0.15 & 370 \\
06/23 12:46 & 530 & 3 & 65 & 2.5 & 2.4 & 65 & 0.45 & 395\\
10/03 00:54 & 450 & 2.2 & 70 & 1.9 & 1.9 & 45 & 0.6 & 390\\
11/04 02:29 & 410 & 2 & 85 & 2.9 & 2.9 & 45 & 0.3 & 360 \\
11/26 05:52 & 440 & 1.9 & 60 & 1.7 & 1.7 & 55 & 0.35 & 375 \\
12/03 04:17 & 480 & 1.8 & 75 & 1.5 & 1.4 & 90 & 0.2 & 370 \\
\hline
\multicolumn{9}{|c|}{\bf 2001}\\
\hline
01/23 10:54 & 610 & 3.1 & 30 & 3.1 & 2.3 & 60 & 0.25 & 425\\
01/31 08:06 & 420 & 2.5 & 45 & 2.8 & 2.5 & 45 & 0.35 & 350\\
03/03 11:24 & 540 & 1.9 & 55 & 1.9 & 1.9 & 50 & 0.45 & 445\\
04/18 00:50 & 550 & 3.5 & 80 & 3 & 3 & 50 & 0.35 & 360 \\
08/12 11:51 & 410 & 3 & 50 & 2.4 & 2.2 & 35 & 0.6 & 320\\
08/27 19:52 & 600 & 2.8 & 45 & 2.8 & 2.8 & 70 & 0.45 & 430\\
10/11 17:06 & 550 & 2.1 & 75 & 2.7 & 2.7 & 70 & 0.1 & 360 \\
10/21 16:46 & 580 & 4.5 & 55 & 2.5 & 2.1 & 60 & 1 & 370 \\
11/19 18:12 & 630 & 3 & 65 & 2 & 1.9 & 70 & 0.35 & 425 \\
\hline
\multicolumn{9}{|c|}{\bf 2002}\\
\hline
03/18 13:21 & 420 & 6 & 40 & 4.5 & 3.2 & 30 & 1.4 & 310 \\
04/23 04:54 & 650 & 3.7 & 35 & 3.4 & 2.2 & 70 & 0.6 & 450 \\
05/18 20:08 & 510 & 4.4 & 50 & 3 & 2.7 & 45 & 0.5 & 330\\
07/17 16:06 & 500 & 2.7 & 30 & 3.2 & 2.3 & 55 & 0.6 & 400\\
08/18 18:47 & 660 & 5.5 & 55 & 3.6 & 3.6 & 45 & 1.8 & 420\\
10/02 22:59 & 540 & 2 & 75 & 2 & 1.9 & 45 & 0.35 & 455\\
\hline
\multicolumn{9}{|c|}{\bf 2003}\\
\hline
06/18 05:19 & 580 & 1.7 & 80 & 1.6 & 1.6 & 90 & 0.2 & 465 \\
11/04 06:32 & 770 & 4.3 & 55 & 4.2 & 2.8 & 60 & 2.3 & 480 \\
11/20 08:05 & 600 & 3.3 & 80 & 3.3 & 1.8 & 80 & 1 & 430\\
\hline
\multicolumn{9}{|c|}{\bf 2004}\\
\hline
04/12 18:26 & 500 & 2.2 & 130 & 2.8 & 2.2 & 55 & 0.5 & 420\\
07/22 10:40 & 480 & 2.3 & 15 & 4 & 2.4 & 65 & 0.15 & 340 \\
08/29 10:16 & 480 & 2.1 & 30 & 1.9 & 1.5 & 40 & 0.5 & 400\\
11/07 18 :28 & 750 & 3 & 60 & 2.3 & 2.3 & 100 & 450 & 1 \\
\hline
\multicolumn{9}{|c|}{\bf 2005}\\
\hline
06/14 18:35 & 540 & 4.5 & 65 & 2.1 & 1.8 & 30 &  0.8 & 435 \\
07/10 03:44 & 520 & 2 &80 & 1.8 & 1.8 & 90 & 0.1 & 350\\
08/01 06:47 & 490 & 1.7 & 85 & 1.9 & 1.9 & 95 & 0.15 & 500\\
09/02 14:24 & 580 & 2.4 & 65 & 2.1 & 2.1 & 60 & 0.6 & 425 \\
09/15 08:52 & 660 & 2.4 & 55 & 2.8 & 2.5 & 60 & 0.4 & 535 \\
\hline
\multicolumn{9}{|c|}{\bf 2006}\\
\hline
09/04 00:12 & 505 & 2.9 & 45 & 2.2 & 1.7 & 45 & 0.7 & 405\\
\hline
\caption{List of 45 shocks from our control sample from 1997 to 2006. The columns list the shock time, shock speed, Alfv\'en Mach number, angle, compression ratios in density and magnetic field; the upstream Alfv{\'e}n speed, $\beta$ and solar wind speed.}
\end{longtable}
\newpage

\end{document}